\PassOptionsToPackage{dvipsnames, table}{xcolor}
\documentclass[AMA,Times1COL]{WileyNJDv5}\usepackage[]{graphicx}
\makeatletter
\def\maxwidth{ %
  \ifdim\Gin@nat@width>\linewidth
    \linewidth
  \else
    \Gin@nat@width
  \fi
}
\makeatother

\definecolor{fgcolor}{rgb}{0.345, 0.345, 0.345}
\newcommand{\hlnum}[1]{\textcolor[rgb]{0.686,0.059,0.569}{#1}}%
\newcommand{\hlsng}[1]{\textcolor[rgb]{0.192,0.494,0.8}{#1}}%
\newcommand{\hlcom}[1]{\textcolor[rgb]{0.678,0.584,0.686}{\textit{#1}}}%
\newcommand{\hlopt}[1]{\textcolor[rgb]{0,0,0}{#1}}%
\newcommand{\hldef}[1]{\textcolor[rgb]{0.345,0.345,0.345}{#1}}%
\newcommand{\hlkwb}[1]{\textcolor[rgb]{0.69,0.353,0.396}{#1}}%
\newcommand{\hlkwc}[1]{\textcolor[rgb]{0.333,0.667,0.333}{#1}}%
\newcommand{\hlkwd}[1]{\textcolor[rgb]{0.737,0.353,0.396}{\textbf{#1}}}%

\usepackage{framed}
\makeatletter
\newenvironment{kframe}{%
 \def\at@end@of@kframe{}%
 \ifinner\ifhmode%
  \def\at@end@of@kframe{\end{minipage}}%
  \begin{minipage}{\columnwidth}%
 \fi\fi%
 \def\FrameCommand##1{\hskip\@totalleftmargin \hskip-\fboxsep
 \colorbox{shadecolor}{##1}\hskip-\fboxsep
     \hskip-\linewidth \hskip-\@totalleftmargin \hskip\columnwidth}%
 \MakeFramed {\advance\hsize-\width
   \@totalleftmargin\z@ \linewidth\hsize
   \@setminipage}}%
 {\par\unskip\endMakeFramed%
 \at@end@of@kframe}
\makeatother

\definecolor{shadecolor}{rgb}{.97, .97, .97}
\definecolor{messagecolor}{rgb}{0, 0, 0}
\definecolor{warningcolor}{rgb}{1, 0, 1}
\definecolor{errorcolor}{rgb}{1, 0, 0}
\newenvironment{knitrout}{}{} 

\usepackage{alltt}
\usepackage{amsmath, amssymb} 
\usepackage{array} 
\usepackage{multirow} 
\usepackage{booktabs} 
\usepackage{nameref} 
\usepackage{orcidlink} 
\definecolor{lightgray}{gray}{0.9} 
\usepackage{pdflscape} 
\usepackage{afterpage} 
\DeclareMathOperator*{\plim}{plim} 

\articletype{Research Article}%
\received{}
\revised{}
\accepted{}
\journal{}
\volume{}
\copyyear{}
\startpage{}
\hypersetup{
  bookmarksopen=true,
  breaklinks=true,
  pdfsubject={},
  pdfkeywords={},
  colorlinks=true,
  linkcolor=black,
  anchorcolor=black,
  citecolor=blue,
  urlcolor=blue,
}

\IfFileExists{upquote.sty}{\usepackage{upquote}}{}
\begin{document}

\title{Combined \textit{P}-value Functions for Compatible Effect Estimation and Hypothesis Testing in Drug Regulation}

\author{Samuel Pawel}

\author{Małgorzata Roos}

\author{Leonhard Held}

\authormark{Pawel \textsc{et al.}}
\titlemark{Combined \textit{P}-value Functions for Compatible Effect Estimation and Hypothesis Testing in Drug Regulation}

\address{\orgdiv{Epidemiology, Biostatistics and Prevention Institute (EBPI), Center for Reproducible Science (CRS)}, \orgname{University of Zurich}, \orgaddress{\state{Zurich}, \country{Switzerland}}}

\corres{Samuel Pawel, Epidemiology, Biostatistics and Prevention Institute, Hirschengraben 84, 8001, Zurich, Switzerland. \email{samuel.pawel@uzh.ch}}

\abstract[Abstract]{The two-trials rule in drug regulation requires statistically
  significant results from two pivotal trials to demonstrate efficacy. However,
  it is unclear how the effect estimates from both trials should be combined to
  quantify the drug effect. Fixed-effect meta-analysis is commonly used but may
  yield confidence intervals that exclude the value of no effect even when the
  two-trials rule is not fulfilled. We systematically address this by recasting
  the two-trials rule and meta-analysis in a unified framework of combined
  \textit{p}-value functions, where they are variants of Wilkinson's and
  Stouffer's combination methods, respectively. This allows us to obtain
  compatible combined \textit{p}-values, effect estimates, and confidence
  intervals, which we derive in closed-form. Additionally, we provide new
  results for Edgington's, Fisher's, Pearson's, and Tippett's \textit{p}-value
  combination methods. When both trials have the same true effect, all methods
  can consistently estimate it, although some show bias. When true effects
  differ, the two-trials rule and Pearson's method are conservative (converging
  to the less extreme effect), Fisher's and Tippett's methods are
  anti-conservative (converging to the more extreme effect), and Edgington's
  method and meta-analysis are balanced (converging to a weighted average).
  Notably, Edgington's confidence intervals asymptotically always include the
  individual trial effects, while meta-analytic confidence intervals shrink to a
  point at the weighted average effect. We conclude that all of these methods
  may be appropriate depending on the estimand of interest. We implement
  combined \textit{p}-value function inference for two trials in the R package
  \texttt{twotrials}, allowing researchers to easily perform compatible
  hypothesis testing and effect estimation.}


\maketitle

\section{Introduction}

The ``two-trials rule'' in drug regulation requires ``\emph{at least two
adequate and well-controlled studies, each convincing on its own}'' for the
demonstration of drug efficacy and subsequent market approval
\citep[p.3]{FDA1998}. This criterion reflects the need for ``substantiation''
and ``replication'' of scientific results \citep[p.8]{FDA2019}, and is typically
implemented by requiring the \textit{p}-values from the two trials to be statistically
significant at the conventional (one-sided) $\alpha = 0.025$ level. However,
this procedure alone does not provide a combined effect estimate nor a
confidence interval (CI), and it has been suggested to pool the estimates with
fixed-effect meta-analysis for this purpose \citep{Fisher1999b, Lu2001,
  Maca2002}. Yet, the meta-analytic CI and point estimate are not always
compatible with the two-trials rule. The meta-analytic CI may exclude the null
value while the two-trials rule is not fulfilled, leading to discrepancies that
are difficult to interpret and communicate.

The results from the two RESPIRE trials \citep{Aksamit2018, DeSoyza2018,
  Chotirmall2018} in Table~\ref{tab:respire} illustrate this phenomenon. While
the \textit{p}-value for the null hypothesis of no effect from RESPIRE 1 is $p =
0.004 < 0.025$, the \textit{p}-value from RESPIRE 2 is $p =
0.144 > 0.025$. Hence, the two-trials rule is not fulfilled at
$\alpha = 0.025$. At the same time, the 95\% CI for the log rate ratio based on
combining the trials' log rate ratio effect estimates with fixed-effect
meta-analysis ranges from $-0.58$ to
$-0.08$ and thus excludes the value of 0.

\begin{table}[!htb]
  \centering
  \caption{Results from the RESPIRE trials regarding the effect of ciprofloxacin
    after 14 days for the treatment of non-cystic fibrosis bronchiectasis
    \citep{Aksamit2018, DeSoyza2018, Chotirmall2018}.}
  \label{tab:respire}
  \begin{tabular}{l c c c}
    \toprule
    & \multicolumn{1}{c}{\textbf{Log rate ratio}} & \textbf{Confidence interval (95\%)} & \multicolumn{1}{c}{\textbf{\textit{P}-value (one-sided)}} \\
    \midrule
    RESPIRE 1 & $-0.49$ & $-0.85$ to $-0.13$ & $0.004$ \\

    RESPIRE 2 & $-0.18$ & $-0.53$ to $\phantom{-}0.16$ & $0.144$ \\
    \midrule
    Meta-analysis & $-0.33$ & $-0.58$ to $-0.08$ & $0.004$ \\
    \bottomrule
  \end{tabular}
\end{table}

A first attempt at resolving the apparent paradox could be to realize that the
confidence level of the CI does not align with the level of the implicit test
underlying the two-trials rule. Since the two-trials rule decision is based on
two independent tests at level $\alpha = 0.025$, the overall test is at level
$\alpha^2 = 0.000625$, thus one could instead take a $(1 - 2 \, \alpha^2)
\times 100\% = 99.875\%$ meta-analytic CI
\citep{Fisher1999, Senn2021}. For the RESPIRE trials, this would lead to a
meta-analytic $99.875\%$ CI from $-0.71$ to $0.05$ which includes the value of 0 and hence
aligns with the two-trials rule decision. However, the level $\alpha = 0.025$ is
arbitrary and it would be desirable to have a CI that is compatible with the
two-trials rule for any level, which is still not the case. For example, for
$\alpha = 0.05$, the two-trials rule is still not fulfilled, while the $(1 - 2
\, \alpha^2) \times 100\% = 99.5\%$ meta-analytic CI from
$-0.66$ to $-0.01$ excludes zero.

Despite the widespread use of the two-trials rule in regulatory decision-making
\citep{Zhang2020}, it remains unclear how point and interval estimation should
be reconciled with it. This paper aims to resolve this issue with a new
approach. The key idea is to look at both the two-trials rule and meta-analysis
from the perspective of \textit{p}-value functions \citep{Bender_etal2005,
  XieSingh2013, Fraser2019, InfangerSchmidt-Trucksass2019, Marschner2024} and
\textit{p}-value combination methods \citep{HedgesOlkin1985, Singh_etal2005,
  cousins2007annotated, Xie_etal2011, Heard2018}. The two-trials rule can be
understood as a combined \textit{p}-value function based on the squared maximum
of two \textit{p}-values \citep{Held2024} which is a special case of Wilkinson's
combination method \citep{Wilkinson1951}, while meta-analysis corresponds to the
combined \textit{p}-value function based on Stouffer's \textit{p}-value
combination method \citep{Stouffer1949} with suitable weights. Both can be used
to obtain combined \textit{p}-values for the null hypothesis of no effect, CIs,
and point estimates. These quantities are compatible in the sense that the
(two-sided) \textit{p}-value for a null value is less than $\alpha$ if and only
if the null value is excluded by the $(1 - \alpha) \times 100\%$ CI, and that
the point estimate is included in the CI at any confidence level $(1 - \alpha)
\in (0,1)$. However, as we will show, the two methods implicitly target
different estimands, which explains their different behaviors, and highlights
the need to choose the method depending on the scientific question and
corresponding estimand of interest. Moreover, the combined \textit{p}-value
function pespective suggests considering alternative \textit{p}-value
combination methods, for example, Edgington's method based on the sum of
\textit{p}-values \citep{Edgington1972} or Fisher's method based on the product
of \textit{p}-values \citep{Fisher1934}. All these \textit{p}-value combination
methods have been studied before in terms of hypothesis testing properties, such
as admissibility or monotonicity \citep{Birnbaum1954, HedgesOlkin1985}. In this
paper, we take an alternative estimation perspective motivated by practical
issues in drug regulation.

This paper is organized as follows: We begin by summarizing the general theory
of combined \textit{p}-value functions (Section~\ref{sec:combinedp}), followed by
investigating combined \textit{p}-value functions based on the two-trials rule
(Section~\ref{sec:2tr}), meta-analysis (Section~\ref{sec:fema}), Tippett's
method (Section~\ref{sec:tippett}), Fisher's and Pearson's methods
(Section~\ref{sec:fisher}), and Edgington's method (Section~\ref{sec:edgington})
in more detail. For each, we derive corresponding point and interval estimates
and investigate their properties. Results from two pairs of clinical trials are
analyzed to illustrate the characteristics of the methods
(Section~\ref{sec:application}). Extensions to more than two trials are
discussed in Section~\ref{sec:extensions}. The paper ends with concluding
discussions, limitations, and an outlook for future research
(Section~\ref{sec:discussion}). Appendix~\ref{app:rpackage} illustrates our R
package \texttt{twotrials} for conducting \textit{p}-value function inference,
while Appendix~\ref{app:technicaldetails} provides additional technical details.

\section{Combined \textit{p}-value functions}
\label{sec:combinedp}
Suppose that two trials yield the effect estimates $\hat{\theta}_{1}$ and
$\hat{\theta}_{2}$ with corresponding standard errors $\sigma_{1}$ and
$\sigma_{2}$, each estimate quantifying the effect of the treatment in the
corresponding trial. Typically, it is reasonable to assume that the effect
estimates (after suitable transformation) are approximately normally distributed
around the trial-specific true effects $\theta_{1}$ and $\theta_{2}$ with
variance equal to their squared standard error, i.e., $\hat{\theta}_{i} \mid
\theta_{i} \sim \mathrm{N}(\theta_{i}, \sigma^{2}_{i})$ for $i \in \{1, 2\}$.
One-sided \textit{p}-values can then be computed by
\begin{equation}
  \label{eq:pnorm}
  p_i(\mu) =
  \begin{cases}
    1-\Phi(Z_i)  & \text{for} ~ H_{1i} \colon \theta_i > \mu ~ \text{(alternative = "greater")} \\
    \Phi(Z_i) &  \text{for} ~ H_{1i} \colon \theta_i < \mu ~ \text{(alternative = "less")} \\
    \end{cases}
\end{equation}
with $z$-values
\begin{equation*}
  Z_i = \frac{\hat \theta_i - \mu}{\sigma_i},
\end{equation*}
cumulative distribution function of the standard normal distribution
$\Phi(\cdot)$, null value $\mu$, and alternative hypothesis $H_{1i}$ chosen
based on the orientation of the effect. For example, if a positive effect
indicates treatment benefit, the alternative "greater" would be chosen. We will
not consider \textit{p}-values with two-sided alternatives here, as the hypotheses
tested in clinical trials usually have a well-defined direction. Moreover,
combined \textit{p}-value functions based on two-sided \textit{p}-values can behave
irregularly, e.g., they can be non-monotone so that the resulting confidence
sets consist of empty or disjoint intervals, which is unintuitive and hard to
communicate \citep{Held_etal2024b}.

A combined \textit{p}-value function $p(\mu)$ is then defined by the function $g$
\begin{equation*}
  p(\mu) = g\left(p_{1}(\mu), p_{2}(\mu)\right),
\end{equation*}
which combines the individual \textit{p}-value functions $p_{1}(\mu)$ and $p_{2}(\mu)$
into a \textit{p}-value function $p(\mu)$, which is a valid \textit{p}-value function in the
sense of having a uniform distribution for a particular $\mu$ if both
$p_{1}(\mu)$ and $p_{2}(\mu)$ are also uniformly distributed for that $\mu$
\citep{XieSingh2013, Held_etal2024b}. A two-sided $(1 - \alpha) \times 100\%$
CI can then be obtained by determining the null values $\mu$
for which the \textit{p}-value function is equal to $\alpha/2$ and $1 - \alpha/2$. The
so-called median estimate is given by the null value $\mu$ for which the
\textit{p}-value function equals 1/2 \citep{Fraser2017}. To obtain these quantities, it
is useful to define a ``combined estimation function''
\begin{equation*}
  \hat{\mu}(a) = \left\{\mu : p(\mu) = a \right\}
\end{equation*}
which is the inverse of the combined \textit{p}-value function. It returns the median
estimate when setting $a = 1/2$, while the limits of a $(1 - \alpha) \times
100\%$ CI are obtained from $a = \alpha/2$ and $a = 1 -
\alpha/2$, respectively. As we will show, combined estimation functions (and
hence the median estimate and any CI) are available in
closed-form for several combined \textit{p}-value functions, including the two-trials
rule and meta-analysis.

\begin{figure}[!htb]
\begin{knitrout}
\definecolor{shadecolor}{rgb}{0.969, 0.969, 0.969}\color{fgcolor}
\includegraphics[width=\maxwidth]{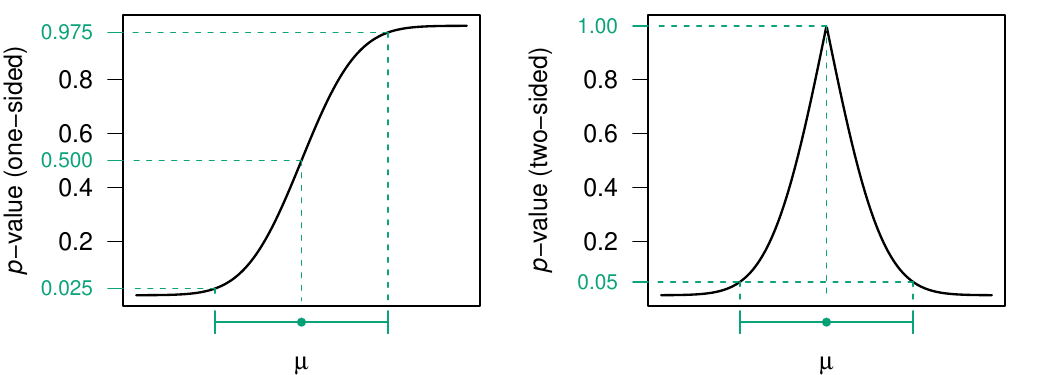} 
\end{knitrout}
\caption{Illustration of a one-sided \textit{p}-value function with alternative =
  "greater" (left plot), corresponding two-sided \textit{p}-value function (right
  plot), and corresponding 95\% CI and median estimate.}
\label{lab:pfunillustration}
\end{figure}

In practice, it is informative to plot the \textit{p}-value function for a range of
null values $\mu$, see the left plot in Figure~\ref{lab:pfunillustration}. For
this purpose, it may also be converted to a two-sided \textit{p}-value function using
the transformation $2\min\{p(\mu), 1 - p(\mu)\}$, known as ``centrality
function'' \citep{XieSingh2013}. Such a two-sided \textit{p}-value function then peaks
at the median estimate, and it can be thresholded at $\alpha$ to conveniently
read off the $(1 - \alpha) \times 100\%$ CI \citep{Held_etal2024b}, see the
right plot in Figure~\ref{lab:pfunillustration}.

When both trials have the same underlying true effect ($\theta_{1} = \theta_{2}
= \theta$), sometimes called ``one population assumption'' or ``homogeneity''
\citep{Shun2005, Zhan2022}, a CI based on a combined \textit{p}-value function
has correct coverage and the median estimate is median unbiased for the true
common effect $\theta$, i.e., the probability of the median estimate being
greater than $\theta$ is equal to the probability of it being smaller than
$\theta$ (see e.g., Xie and Singh\citep{XieSingh2013}). However, it is unclear
how other operating characteristics (e.g., mean bias or CI width) behave for
different combined \textit{p}-value functions $g$, and how they behave when the
true effects are not the same ($\theta_{1} \neq \theta_{2}$), known as ``two
populations assumption'' or ``heterogeneity'' \citep{Shun2005, Zhan2022}. In the
following, we will investigate this in detail for the two-trials rule,
meta-analysis, and four other types of combined \textit{p}-value functions. As
these investigations are somewhat technical, readers may choose to look only at
the summary in Table~\ref{tab:summaries} and then jump directly to the
applications in Section~\ref{sec:application}.

\afterpage{
\begin{landscape}
\begingroup
\renewcommand{\arraystretch}{1.3} 
\begin{table}[!htb]
  \centering
  \caption{Summary of combined \textit{p}-value functions and corresponding estimation
    functions. All are based on the alternative ``greater''. The median estimate
    is obtained from setting $a = 1/2$, while the limits of a $(1 - \alpha)
    \times 100\%$ confidence interval (CI) are obtained from $a = \alpha/2$ and
    $a = 1 - \alpha/2$, respectively.}
  \label{tab:summaries}
  \rowcolors{1}{}{gray!15}
  \resizebox{1\linewidth}{!}{%
  \begin{tabular}{p{0.11\linewidth}   p{0.22\linewidth}  p{0.22\linewidth}  >{\footnotesize}p{0.26\linewidth}}
    \toprule
    \multicolumn{1}{c}{\textbf{Method}} &
    \multicolumn{1}{c}{\textbf{Combined \textit{p}-value function}} &
    \multicolumn{1}{c}{\textbf{Combined estimation function}} &
    \multicolumn{1}{c}{\textbf{Properties}} \\
    \midrule

    \textbf{Two-trials rule} \newline Maximum \textit{p}-value, special case of Wilkinson's method, Section~\ref{sec:2tr} &
    \vfill
    \centering
    \( p_{\text{2TR}}(\mu) = \max\{p_{1}(\mu), p_{2}(\mu)\}^{2} \)
    \newline ~ \newline
    R function \texttt{twotrials::p2TR}
    &
    \vfill
    \centering
    \( \hat{\mu}_{\text{2TR}}(a) =
    \min\{\hat{\theta}_1 + \sigma_1 \, z_{\sqrt{a}}, \hat{\theta}_2 + \sigma_2 \, z_{\sqrt{a}}\} \)
    \newline ~ \newline
    R function \texttt{twotrials::mu2TR}
    &
    -- Targets least extreme true effect (conservative) \newline
    -- Mean-biased when trials have the same true effects \newline
    -- CI shrinks to point with decreasing standard errors \newline
    -- Median estimate not equal to observed effect estimates when the same estimates in both trials \newline
    -- Median estimate standard error can be larger than trial standard errors
    \\

    \textbf{Fixed-effect \newline meta-analysis} \newline Weighted Stouffer's method, inverse-normal method, Section~\ref{sec:fema} &
    \vfill
    \centering
    \(p_{\text{MA}}(\mu) = 1 - \Phi(Z_{\text{MA}}) \)
    \newline ~ \newline
    with~\(Z_{\text{MA}} = \frac{\Phi^{-1}\{1-p_1(\mu)\}/\sigma_{1} + \Phi^{-1}\{1-p_2(\mu)\}/\sigma_{2}}{\sqrt{1/\sigma_{1}^{2} + 1/\sigma_{2}^{2}}}\)
    \newline ~ \newline
    R function \texttt{twotrials::pMA}
    &
    \vfill
    \centering
    \(\hat{\mu}_{\text{MA}}(a) = \hat{\theta}_{\text{MA}} + \sigma_{\text{MA}} \, z_{a}\)
    \newline \newline with ~
    \parbox{3cm}{\(\begin{aligned}
      \sigma_{\text{MA}}^{2} &= 1/(1/\sigma_{1}^{2} + 1/\sigma_{2}^{2}) \\
      \hat{\theta}_{\text{MA}} &= (\hat{\theta}_{1}/\sigma_{1}^{2} + \hat{\theta}_{2}/\sigma_{2}^{2}) \, \sigma_{\text{MA}}^{2}
      \end{aligned}\)}
    \newline ~ \newline
    R function \texttt{twotrials::muMA}
    &
    -- Targets weighted average effect (inverse squared standard error weights) \newline
    -- Mean-unbiased when the same true effects \newline
    -- CI shrinks to point with decreasing standard errors \newline
    -- Median estimate equals observed effect estimates when the same estimates in both trials \newline
    -- Median estimate standard error cannot be larger than trial standard errors
    \\

    \textbf{Tippett's method} \newline Minimum \textit{p}-value, special case of Wilkinson's method, Section~\ref{sec:tippett} &
    \vfill
    \centering
    \( p_{\text{T}}(\mu) = 1 - (1 - \min\{p_{1}(\mu), p_{2}(\mu)\})^{2} \)
    \newline ~ \newline
    R function \texttt{twotrials::pTippett}
    &
    \vfill
    \centering
    \( \hat{\mu}_{\text{T}}(a) =
    \max\{\hat{\theta}_1 - \sigma_1 \, z_{\sqrt{1 - a}}, \hat{\theta}_2 - \sigma_2 \, z_{\sqrt{1 - a}}\} \)
    \newline ~ \newline
    R function \texttt{twotrials::muTippett}
    &
    -- Targets most extreme true effect (anti-conservative) \newline
    -- Mean-biased when the same true effects \newline
    -- CI shrinks to point with decreasing standard errors \newline
    -- Median estimate not equal to observed effect estimates when the same estimates in both trials \newline
    -- Median estimate standard error can be larger than trial standard errors
    \\

    \textbf{Fisher's method} \newline Product of \textit{p}-values, Section~\ref{sec:fisher} &
    \vfill
    \centering
    \( p_{\text{F}}(\mu) = 1 - \Pr(\chi^{2}_{4} \leq F) \)
    \newline ~ \newline
    with
    \(F = -2 [\log\{p_{1}(\mu)\} + \log \{p_{2}(\mu)\}]\)
    \newline ~ \newline
    R function \texttt{twotrials::pFisher}
    &
    \vfill
    \centering
    $\hat{\mu}_{\text{F}}(a)$ not analytically available
    \newline ~ \newline
    R function \texttt{twotrials::muFisher}
    &
    -- Targets most extreme true effect (anti-conservative) \newline
    -- CI shrinks to point with decreasing standard errors \newline
    -- Median estimate not equal to observed effect estimates when the same estimates in both trials \newline
    -- Median estimate standard error can be larger than trial standard errors
    \\

    \textbf{Pearson's method} \newline Product of $1 - p$-values, Section~\ref{sec:fisher} &
    \vfill
    \centering
    \( p_{\text{P}}(\mu) = \Pr(\chi^{2}_{4} \leq K) \)
    \newline ~ \newline
    with~\(K = -2 [\log\{1 - p_{1}(\mu)\} + \log \{1 - p_{2}(\mu)\}]\)
    \newline ~ \newline
    R function \texttt{twotrials::pPearson}
    &
    \vfill
    \centering
    $\hat{\mu}_{\text{P}}(a)$ not analytically available
    \newline ~ \newline
    R function \texttt{twotrials::muPearson}
    &
    -- Targets least extreme true effect (conservative) \newline
    -- CI shrinks to point with decreasing standard errors \newline
    -- Median estimate not equal to observed effect estimates when the same estimates in both trials \newline
    -- Median estimate standard error can be larger than trial standard errors
    \\

    \textbf{Edgington's method} \newline Sum of \textit{p}-values, \newline Section~\ref{sec:edgington} &
    \vfill
    \parbox{3cm}{\(\begin{aligned}
      p_{\text{E}}(\mu) =
      \begin{cases}
        E^{2}/2 & \text{if} ~ 0 \leq E \leq 1 \\
        1 - (2 - E)^{2}/2 & \text{if} ~ 1 < E \leq 2 \\
      \end{cases}
    \end{aligned}\)}
    \newline ~ \newline
    with
    \(E = p_1(\mu) + p_2(\mu)\)
    \newline ~ \newline
    R function \texttt{twotrials::pEdgington}
    &
    \vfill
    \centering
    Median estimate analytically available
    \(\hat{\mu}_{\text{E}}(a = 1/2) = \dfrac{\hat{\theta}_1/\sigma_1 + \hat{\theta}_2/\sigma_2}{1/\sigma_1 + 1/\sigma_2}\)
    \newline ~ \newline
    $\hat{\mu}_{\text{E}}(a)$ not analytically available for \(a \neq 1/2\)
    \newline
     R function \texttt{twotrials::muEdgington}
    &
    -- Targets weighted average effect (inverse standard error weights) \newline
    -- Mean-unbiased when the same true effects \newline
    -- CI asymptotically always includes both true effects (only shrinks to point when both are equal) \newline
    -- Median estimate equals observed effect estimates when the same estimates in both trials \newline
    -- Median estimate standard error can be larger than trial standard errors
    \\
    \bottomrule
  \end{tabular}
  }
\end{table}
\endgroup
\end{landscape}
}

\subsection{The two-trials rule (maximum method)}
\label{sec:2tr}
The two-trials rule is fulfilled if $\max\{p_{1}, p_{2}\} \leq \alpha$, or
equivalently if
\begin{align}
  \label{eq:2trials}
  p_{\text{2TR}}(\mu) = \max\left\{p_{1}(\mu), p_{2}(\mu)\right\}^{2} \leq \alpha^{2}.
\end{align}
The formulation using the squared maximum~\eqref{eq:2trials} may be preferable
because $p_{\text{2TR}}(\mu)$ is a valid \textit{p}-value, i.e., it has a uniform
distribution if both $p_{1}(\mu)$ and $p_{2}(\mu)$ are also uniformly
distributed for a particular $\mu$ \citep{Held2024}. The combined
\textit{p}-value function~\eqref{eq:2trials} is also a special case of
Wilkinson's \textit{p}-value combination method based on the $r$th smallest out of $k$
\textit{p}-values with $r = k = 2$ \citep{Wilkinson1951}. This relationship can be used
to generalize the two-trials rule to different settings while preserving type I
error control at level $\alpha^2$, for example, settings with three rather than
two trials \citep{Rosenkranz2023}. We will discuss such extensions to more than
two trials in Section~\ref{sec:extensions} and focus first on effect estimation
for two trials.

\subsubsection{Effect estimation}
In order to obtain a CI and a point estimate based on the
two-trials rule, we can equate the combined \textit{p}-value
function~\eqref{eq:2trials} to some value $a \in (0, 1)$ and solve for the null
value~$\mu$. This leads to the combined estimation function
\begin{equation}
  \label{eq:2trialsest}
  \hat{\mu}_{\text{2TR}}(a) =
  \begin{cases}
    \min\{\hat{\theta}_1 + \sigma_1 \, z_{\sqrt{a}}, \hat{\theta}_2 + \sigma_2 \, z_{\sqrt{a}}\} & \text{for alternative = "greater"} \\
    \max\{\hat{\theta}_1 - \sigma_1 \, z_{\sqrt{a}},  \hat{\theta}_2 - \sigma_2 \, z_{\sqrt{a}}\} & \text{for alternative = "less"}
  \end{cases}
\end{equation}
with $z_{q}$ the $q \times 100\%$ quantile of the standard normal distribution.
For $a = 1/2$ the median estimate is obtained, while the limits of an $(1
- \alpha) \times 100\%$ CI can be obtained from $a = \alpha/2$
and $a = 1 - \alpha/2$.

\begin{figure}[!htb]
\begin{knitrout}
\definecolor{shadecolor}{rgb}{0.969, 0.969, 0.969}\color{fgcolor}
\includegraphics[width=\maxwidth]{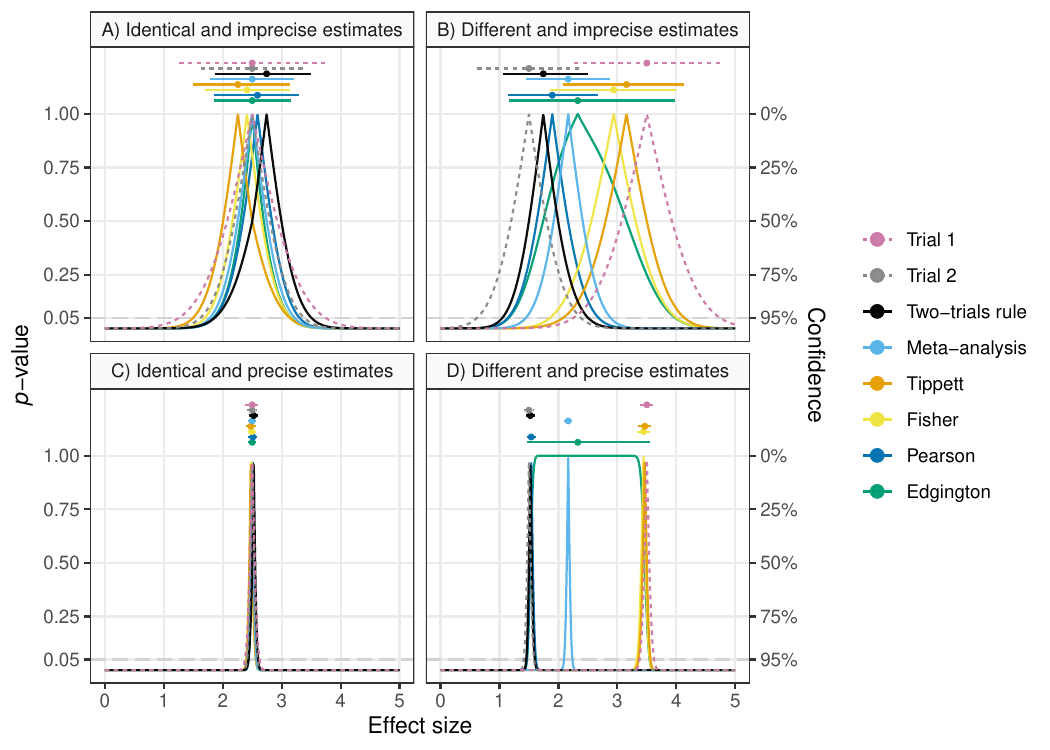} 
\end{knitrout}
\caption{Four hypothetical pairs of effect estimates and standard errors from
  two trials. The standard errors are assumed to be of the form $\sigma =
  \sqrt{2/n}$. The sample size $n$ in trial 1 is set to 5
  (imprecise) or 500 (precise), and in trial 2 to twice the sample
  size of trial 1. The two-sided \textit{p}-value functions of the individual
  trials (dashed lines), and the combined \textit{p}-value functions (solid
  lines) based on the two-trials rule, fixed-effect meta-analysis, Tippett's,
  Fisher's, Pearson's, and Edgington's methods are shown along with the
  corresponding 95\%~CIs and median estimates (top). All \textit{p}-values are
  based on the alternative "greater" and then converted to two-sided
  \textit{p}-values via the centrality function $2\min\{p, 1 - p\}$.}
\label{fig:extremeExamples}
\end{figure}

Now assume that the standard errors of both trials are the same ($\sigma_{1} =
\sigma_{2} = \sigma$) and the alternative is "greater". The median estimate is
then
\begin{equation}
  \label{eq:2trpe}
  \hat{\mu}_{\text{2TR}}(1/2) = \min\{\hat{\theta}_{1}, \hat{\theta}_{2}\} + \sigma \, \underbrace{z_{\sqrt{1/2}}}_{0.54}
\end{equation}
and the 95\% CI is given by
\begin{equation}
  \label{eq:2trci}
  \bigg[
    \min\{\hat{\theta}_{1}, \hat{\theta}_{2}\} + \sigma \, \underbrace{z_{\sqrt{0.025}}}_{-1},\,
    \min\{\hat{\theta}_{1}, \hat{\theta}_{2}\} + \sigma \, \underbrace{z_{\sqrt{0.975}}}_{2.24}
  \bigg].
\end{equation}
Both seem counterintuitive. For instance, if the trial effect estimates are the
same ($\hat{\theta}_{1} = \hat{\theta}_{2} = \hat{\theta}$), the median
estimate~\eqref{eq:2trpe} is shifted away from the observed estimate by $\sigma
\times z_{\sqrt{1/2}} \approx \sigma \times 0.54$,
and also the CI~\eqref{eq:2trci} is not centered around it. This is illustrated
in Figure~\ref{fig:extremeExamples} (panels A and C), where the hypothetical
trial effect estimates are identical, but the median estimates based on the
two-trials rule (black) are larger. Moreover, the CI~\eqref{eq:2trci} is skewed
in the sense that the distance between the upper limit and the median estimate
is larger than the distance between the lower limit and the median estimate
although the estimates are the same.

While this CI has correct coverage and the median estimate is
median unbiased we may look at other operating characteristics. The expectation
of the median estimate~\eqref{eq:2trpe} can be derived to be
\begin{equation}
  \label{eq:exp2tr}
  \mathrm{E}\left[\hat{\mu}_{\text{2TR}}(1/2)\right] =
  \theta_{1}\, \Phi\left(\frac{\theta_{2} - \theta_{1}}{\sqrt{2}\,\sigma}\right) +
  \theta_{2}\, \Phi\left(\frac{\theta_{1} - \theta_{2}}{\sqrt{2}\,\sigma}\right) +
  \sigma\, \left\{
  z_{\sqrt{1/2}} -
  \sqrt{2}\,\phi\, \left(\frac{\theta_{2} - \theta_{1}}{\sqrt{2}\,\sigma}\right)
  \right\}
\end{equation}
where $\phi(\cdot)$ denotes the density function of the standard normal
distribution, see Appendix~\ref{app:expectation} for details. If the true
effects from the two trials coincide ($\theta_{1} = \theta_{2} = \theta$) the
first two terms of the expectation~\eqref{eq:exp2tr} reduce to the common effect
$\theta$, whereas the last term reduces to $\sigma \times (z_{\sqrt{1/2}} -
1/\sqrt{\pi}) \approx \sigma \times -0.019$. Hence, the median estimate with~\eqref{eq:2trpe} is negatively biased,
yet the bias vanishes as the standard error decreases. In a similar way, one can
show that the median estimate for the alternative "less" is positively biased,
so the median estimate from the two-trials rule exhibits a conservative bias in
both cases.

Another interesting operating characteristic is the standard error of the median
estimate. An intuitively desirable property is that the standard error of a
combined estimate should not be larger than either of the trials' standard
errors. Assuming again that the true trial effects and standard errors coincide,
the two-trials rule satisfies this, as the standard error takes the simple form
\begin{equation}
  \label{eq:se2tr}
  \sigma_{\text{2TR}} =
  \sigma \sqrt{1 - 1/\pi}
  \approx \sigma \times 0.83,
\end{equation}
so is approximately $17\%$ smaller than the standard errors of each individual
trial. However, this is no longer the case when the trial standard errors
differ, where the standard error of the two-trial rule also depends on the true
trial effects. See Appendix~\ref{app:ses} for more details, also on standard
errors based on other combined \textit{p}-value functions.

\subsubsection{Asymptotics}
Suppose now that the sample size of the trials increases and in turn the
standard errors of the effect estimates decrease toward zero. The combined
estimation function~\eqref{eq:2trialsest} then converges to
\begin{align*}
  \plim_{\sigma_1, \sigma_2 \downarrow 0} \hat{\mu}_{\text{2TR}}(a) =
    \begin{cases}
    \min\{\theta_1, \theta_2\} & \text{for alternative = "greater"} \\
    \max\{\theta_1, \theta_2\} & \text{for alternative = "less"},
  \end{cases}
\end{align*}
see Appendix~\ref{app:asymptotics} for details. Hence, the median estimate ($a =
1/2$) and any CI limit ($a \neq 1/2$) approach
$\min\{\hat{\theta}_1, \hat{\theta}_2\}$ or $\max\{\hat{\theta}_1,
\hat{\theta}_2\}$, depending on the alternative hypothesis. This means that the
CI shrinks to the more conservative of the two effects, while
in case they coincide ($\theta_1 = \theta_2 = \theta$) it shrinks to the common
effect $\theta$. Both scenarios are illustrated in
Figure~\ref{fig:extremeExamples}: In case of very small standard errors and
different effect estimates (panel D), the CI based on the
two-trials rule (black) is tightly concentrated around the smaller effect
estimate, while for identical effect estimates (panel C), it is tightly
concentrated around the common effect estimate.

\subsection{Fixed-effect meta-analysis (Stouffer's method)}
\label{sec:fema}
We will now compare the \textit{p}-value function from the two-trials rule with its
meta-analysis counterpart. The combined \textit{p}-value based on fixed-effect
meta-analysis is given by
\begin{align}
  \label{eq:pMeta}
  p_{\text{MA}}(\mu) =
    \begin{cases}
    1 - \Phi(Z_{\text{MA}}) & \text{for alternative = "greater"} \\
    \Phi(Z_{\text{MA}}) & \text{for alternative = "less"}
  \end{cases}
\end{align}
with
\begin{align}
  Z_{\text{MA}}
  = \frac{Z_{1}/\sigma_{1} + Z_{2}/\sigma_{2}}{\sqrt{1/\sigma_{1}^{2} + 1/\sigma_{2}^{2}}}
  = \frac{\hat{\theta}_{\text{MA}} - \mu}{\sigma_{\text{MA}}} \label{eq:zMeta}
\end{align}
where
\begin{align}
  \label{eq:tMA}
 \hat{\theta}_{\text{MA}} = \frac{\hat{\theta}_{1}/\sigma_{1}^{2} +
    \hat{\theta}_{2}/\sigma_{2}^{2}}{1/\sigma_1^2 + 1/\sigma_2^2}
\end{align}
and
\begin{align}
  \label{eq:seMA}
  \sigma_{\text{MA}} = \frac{1}{\sqrt{1/\sigma_{1}^{2} + 1/\sigma_{2}^{2}}}.
\end{align}
The first equation in~\eqref{eq:zMeta} represents Stouffer's \textit{p}-value
combination method (after transforming \textit{p}-values to $z$-values) using
inverse standard errors as weights \citep{cousins2007annotated}, whereas the
second equation in~\eqref{eq:zMeta} shows the corresponding representation via
the meta-analytically pooled estimate~\eqref{eq:tMA} and standard
error~\eqref{eq:seMA} \citep{Borenstein2010}. While meta-analytic pooling could
be extended to the random-effects model, this is typically not desired with only
two studies for three reasons. First, the interest is in the true effects
underlying the studies. Second, random-effects variance estimation is unreliable
with only two studies. Finally, even if there is effect heterogeneity,
fixed-effect meta-analysis is a valid procedure which estimates a well-defined
average true effect \citep{Rice_etal2018}.

\subsubsection{Effect estimation}
To obtain meta-analytic CIs and point estimates we can also
equate the \textit{p}-value function~\eqref{eq:pMeta} to $a$ and solve for $\mu$. This
leads to the combined estimation function
\begin{align*}
  \hat{\mu}_{\text{MA}}(a) =
  \begin{cases}
   \hat{\theta}_{\text{MA}} + \sigma_{\text{MA}} \, z_{a}  & \text{for alternative = "greater"} \\
   \hat{\theta}_{\text{MA}} - \sigma_{\text{MA}} \, z_{a}  & \text{for alternative = "less"}.
  \end{cases}
\end{align*}
When $a = 1/2$ we obtain $\hat{\theta}_{\text{MA}}$ as the median estimate,
while $a = \alpha/2$ and $a = 1 - \alpha/2$ give the limits of the $(1 -\alpha)
\times 100\%$ CI corresponding to the usual fixed-effect
meta-analytic Wald CI.

The standard error~\eqref{eq:seMA} of the meta-analytic median estimate has two
desirable properties: First, it is never larger than either of trials' standard
errors ($\sigma_{\text{MA}} \leq \min\{\sigma_1, \sigma_2\}$). Second, under
effect homogeneity, the standard error is the smallest among all unbiased
estimators of the common effect \citep{HedgesOlkin1985}. Both properties do not
hold for the two-trials rule and the other \textit{p}-value combination methods
discussed below.

\subsubsection{Asymptotics}
Since the meta-analytic combined estimation function is a linear combination of
normally distributed effect estimates, its distribution is also normal and given
by
\begin{align*}
  \hat{\mu}_{\text{MA}}(a) \sim
  \mathrm{N}\left(\frac{\theta_{1}}{1 + c} + \frac{\theta_{2}}{1 + 1/c} -
  \frac{z_a}{\sqrt{1/\sigma^2_1 + 1/\sigma^2_2}}, \,
  \frac{1}{1/\sigma^2_1 + 1/\sigma^2_2} \right)
\end{align*}
for the alternative ``greater'' and with variance ratio $c =
\sigma^{2}_{1}/\sigma^{2}_{2}$. For the alternative ``less'', the minus in the
mean has to be replaced with a plus. The median estimate ($a = 1/2$) hence
targets the weighted average of the true effects
\begin{align*}
  \frac{\theta_{1}}{1 + c} + \frac{\theta_{2}}{1 + 1/c}
\end{align*}
while the meta-analytic CIs becomes increasingly concentrated
around the weighted average with decreasing standard errors, provided the
relative variance $c$ stays constant.

Meta-analysis thus shows a less conservative asymptotic behavior than the
two-trials rule in the sense that a more extreme effect can compensate for a
less extreme one, whereas the two-trials rule would converge to the less extreme
of the two effects. Figure~\ref{fig:extremeExamples} illustrates this asymptotic
behavior: In case both estimates are identical and the standard errors very
small (panel C), the meta-analytic CI is concentrated around the trials
estimate, while in case of different estimates the CI concentrates somewhere in
between (panel D). Since in this example the relative variance is $c = 2$, the
weighted average is slightly closer to the estimate from trial 2.

\section{Other \textit{p}-value combination methods}
While the two-trials rule and meta-analysis are the most commonly used \textit{p}-value
combination methods in practice, several other combination methods exist
\citep{HedgesOlkin1985}. In this section, we examine Tippett's, Fisher's,
Pearson's, and Edgington's methods, which can also be used to obtain combined
effect estimates, CIs, and \textit{p}-values. Although these methods are not standard
in drug regulation, they may have useful properties in certain settings, as we
will demonstrate in the following.

\subsection{Tippett's (minimum) method}
\label{sec:tippett}
 The combined \textit{p}-value from Tippett's method \citep{Tippett1931} is based on
 the minimum of the two \textit{p}-values and given by
\begin{equation*}
  p_{\text{T}}(\mu) = 1 - (1 - \min\{p_{1}(\mu), p_{2}(\mu)\})^{2}.
\end{equation*}
It is closely related to the two-trials rule in the sense that the combined
\textit{p}-value based on the alternative ``greater'' from Tippett's method is the same
as one minus the combined \textit{p}-value based on the alternative ``less'' from the
two-trials rule, and vice versa \citep{Held_etal2024b}. Similarly, Tippett's
method is a special case of Wilkinson's method based on the $r = 1$ smallest out
of $k = 2$ \textit{p}-values.

\subsubsection{Effect estimation}
Following a similar approach as with the two-trials rule, CIs
and point estimates based on Tippett's method can be obtained in closed-form
with the combined estimation function
\begin{equation}
  \label{eq:tippest}
  \hat{\mu}_{\text{T}}(a) =
  \begin{cases}
    \max\{\hat{\theta}_1 - \sigma_1 \, z_{\sqrt{1 - a}}, \hat{\theta}_2 - \sigma_2 \, z_{\sqrt{1 - a}}\} & \text{for alternative = "greater"} \\
    \min\{\hat{\theta}_1 + \sigma_1 \, z_{\sqrt{1 - a}},  \hat{\theta}_2 + \sigma_2 \, z_{\sqrt{1 - a}}\} & \text{for alternative = "less"}.
  \end{cases}
\end{equation}
The similarity to the two-trials rule is again visible as~\eqref{eq:tippest}
looks similar to the estimation function from the two-trials
rule~\eqref{eq:2trialsest} with the minimum and maximum flipped and using
different normal quantiles. In particular, the same median estimates are
obtained (i.e., $\hat{\mu}_{\text{2TR}}(1/2) = \hat{\mu}_{\text{T}}(1/2)$) if
opposite alternatives are specified.

We can see that when the observed effect estimates are the same ($\hat{\theta}_1
= \hat{\theta}_2 = \hat{\theta}$), the median estimate ($a = 1/2$) based on
Tippett's method is not equal to $\hat{\theta}$ but shifted from it, as the
two-trials rule (see panels A and C in Figure~\ref{fig:extremeExamples} for an
illustration). Similarly, CIs obtained from Tippett's method are typically
skewed in the sense that the distances between the point estimate and the upper
and lower limits are not the same.

\subsubsection{Asymptotics}
It can be shown that as the standard errors $\sigma_1$ and $\sigma_2$ decrease,
the combined estimation function~\eqref{eq:tippest} converges to
\begin{align*}
  \plim_{\sigma_1, \sigma_2 \downarrow 0} \hat{\mu}_{\text{T}}(a) =
    \begin{cases}
    \max\{\theta_1, \theta_2\} & \text{for alternative = "greater"} \\
    \min\{\theta_1, \theta_2\} & \text{for alternative = "less"},
  \end{cases}
\end{align*}
that is, the more extreme of the two effects, see Appendix~\ref{app:asymptotics}
for details. In contrast to the two-trials rule, Tippett's method is hence
anti-conservative. This is illustrated in panel D of
Figure~\ref{fig:extremeExamples} where Tippett's CI is tightly concentrated
around the larger effect estimate.

\subsection{Fisher's and Pearson's (product) methods}
\label{sec:fisher}
Pearson's and Fisher's combination method are two closely related \textit{p}-value
combination methods, that are based on the product of \textit{p}-values, or
equivalently, the sum of the log \textit{p}-values. Fisher's method has been proposed
for combining \textit{p}-values from clinical trials \citep{Fisher1999, Rosenkranz2002,
  Shun2005}, however, using the associated \textit{p}-value function for effect
estimation in a regulatory trials setting has remained unexplored.

The combined \textit{p}-value function based on Fisher's method \citep{Fisher1934} is
given by
\begin{equation}
  \label{eq:fisher}
  p_{\text{F}}(\mu) =
  1 - \Pr\left(\chi^{2}_{4} \leq -2 [\log\{p_{1}(\mu)\} + \log \{p_{2}(\mu)\}]\right)
\end{equation}
while the combined \textit{p}-value function based on Pearson's method
\citep{Pearson1933} is given by
\begin{equation}
  \label{eq:pearson}
  p_{\text{P}}(\mu) =
  \Pr\left(\chi^{2}_{4} \leq -2 [\log\{1 - p_{1}(\mu)\} + \log \{1 - p_{2}(\mu)\}]\right).
\end{equation}
Pearson \citep{Pearson1934} proposed also another method based on the maximum of
the test statistics underlying the \textit{p}-values~\eqref{eq:fisher}
and~\eqref{eq:pearson}, but we will not consider this method here as its test
statistic does not have an exact null distribution \citep{Owen2009}. As with the
two-trials rule and Tippett's method, the combined \textit{p}-value functions of
Fisher's and Pearson's methods are related in the sense that the
\textit{p}-value function based on Fisher's method and the alternative
hypothesis ``greater'' is the same as one minus the \textit{p}-value function
based on Pearson's method and the alternative ``less'', and vice versa
\citep{Held_etal2024b}. As we will show in the following, Fisher's method also
acts in a similar anti-conservative way as Tippett's method, while Pearson's
method acts in a similar conservative way as the two-trials rule.

\subsubsection{Effect estimation}
CIs and point estimates based on Fisher's and Pearson's methods
can in general not be obtained in closed-form but require numerical
root-finding. However,
a special case where a closed-form solution is available is when the effect
estimates and standard errors are the same in both trials ($\hat{\theta}_1 =
\hat{\theta}_2 = \hat{\theta}$ and $\sigma_1 = \sigma_2 = \sigma$). While this
is unrealistic in practice, it serves as an important soundness check to
investigate whether the methods produce reasonable estimates in the situation of
identical trial results. In this case, we obtain the following closed-form
combined estimation function for Fisher's method
\begin{equation}
  \label{eq:fisherest}
  \hat{\mu}_{\text{F}}(a) =
  \begin{cases}
    \hat{\theta} + \sigma \, z_{\exp\{-\chi^{2}_{4}(1 - a)/4\}} &  \text{for alternative = "greater"} \\
    \hat{\theta} - \sigma \, z_{\exp\{-\chi^{2}_{4}(1 - a)/4\}} &  \text{for alternative = "less"} \\
  \end{cases}
\end{equation}
and for Pearson's method
\begin{equation}
  \label{eq:pearsonest}
  \hat{\mu}_{\text{P}}(a) =
  \begin{cases}
    \hat{\theta} - \sigma \, z_{\exp\{-\chi^{2}_{4}(a)/4\}} &  \text{for alternative = "greater"} \\
    \hat{\theta} + \sigma \, z_{\exp\{-\chi^{2}_{4}(a)/4\}} &  \text{for alternative = "less"} \\
  \end{cases}
\end{equation}
with $\chi^{2}_{4}(a)$ the $a\times 100\%$ quantile of the chi-squared
distribution with four degrees of freedom. Importantly, the median estimates ($a
= 1/2$) from both methods do not equal the observed estimate $\hat{\theta}$ but
are shifted away from it by $z_{\exp\{-\chi^{2}_{4}(1/2)/4\}} \approx
-0.17$ standard errors
$\sigma$, similar to the two-trials rule and Tippett's method. Another
similarity is that the CI is skewed since the distance between the lower and
upper limits to the point estimate is not the same.

\subsubsection{Asymptotics}
To understand the asymptotic behavior of Fisher's and Pearson's method, we may
again examine their combined estimation functions for decreasing standard
errors. When the true effects are equal ($\theta_1 = \theta_2 = \theta$), both
Fisher's and Pearson's median estimates will converge toward it, which is clear
from the theory of \textit{p}-value functions but can also be informally seen
from~\eqref{eq:fisherest} and~\eqref{eq:pearsonest} shrinking toward the common
effect estimate for a decreasing standard error. On the other hand, when the
true effects are unequal, it can then be shown that
\begin{align*}
  \plim_{\sigma_1, \sigma_2 \downarrow 0} \hat{\mu}_{\text{F}}(a) =
    \begin{cases}
    \max\{\theta_1, \theta_2\} & \text{for alternative = "greater"} \\
    \min\{\theta_1, \theta_2\} & \text{for alternative = "less"},
  \end{cases}
\end{align*}
and
\begin{align*}
  \plim_{\sigma_1, \sigma_2 \downarrow 0} \hat{\mu}_{\text{P}}(a) =
    \begin{cases}
    \min\{\theta_1, \theta_2\} & \text{for alternative = "greater"} \\
    \max\{\theta_1, \theta_2\} & \text{for alternative = "less"},
  \end{cases}
\end{align*}
see Appendix~\ref{app:approxmu} for details. This means that the combined
estimation functions converge toward the more extreme effect for Fisher's method
(e.g., the maximum of two positively oriented effects), and the less extreme
effect for Pearson's method (e.g., the minimum of two positively oriented
effects). The behavior is similar to Tippett's method and the two-trials rule
where one method acts anti-conservative (Fisher and Tippett's methods), while
the other methods acts conservative (Pearson's method and the two-trials rule).
However, the examples in panels B and D of Figure~\ref{fig:extremeExamples}
suggest that in finite samples, Fisher's and Pearson's method remain closer to
the weighted average compared to Tippett's method and the two-trials rule.

\subsection{Edgington's (sum) method}
\label{sec:edgington}
Edgington's method based on the sum of \textit{p}-values \citep{Edgington1972,
  Held_etal2024} is yet another \textit{p}-value combination method that can be
used for obtaining a combined \textit{p}-value function, and the last method
that we will consider in this paper. It is given by
\begin{equation}
  \label{eq:edgington}
  p_{\text{E}}(\mu) =
  \begin{cases}
    E^{2}/2 & \text{if} ~ 0 \leq E \leq 1 \\
    1 - (2 - E)^{2}/2 & \text{if} ~ 1 < E \leq 2 \\
  \end{cases}
\end{equation}
with $E = p_{1}(\mu) + p_{2}(\mu)$. An attractive feature is that two-sided CIs
based on Edgington's method are orientation invariant, which is not the case for
the other combined \textit{p}-value functions considered so far. That is, CIs based on
Edgington's method do not depend on the orientation of the underlying one-sided
\textit{p}-values, so the same CI is obtained regardless whether one uses \textit{p}-values
with the alternative "greater" or "less" \citep{Held_etal2024b}. Edgington's
method has previously been used in meta-analysis \citep{Held_etal2024b}, to
synthesize \textit{p}-values from original and replication studies
\citep{Held_etal2024}, and suggested as an alternative for the two-trials rule
\citep{Held2024}. However, its estimation properties in the context of two
trials remain unexplored.

\subsubsection{Effect estimation}
The median estimate based on Edgington's method has an intuitive interpretation
as the null value $\mu$ for which the sum of the \textit{p}-values is one. It can be
obtained in closed-form by
\begin{equation}
  \label{eq:edgingtonpointest}
  \hat{\mu}_{\text{E}}(1/2) =
  \frac{\hat{\theta}_1/\sigma_1 + \hat{\theta}_2/\sigma_2}{1/\sigma_1 + 1/\sigma_2}
\end{equation}
so is a weighted average of the two effect estimates, as the meta-analytic point
estimate~\eqref{eq:tMA}. However, the weights from Edgington's method are equal
to the inverse standard errors, whereas the weights from meta-analysis are equal
to the inverse squared standard errors. Thus, Edgington's method gives more
weight to smaller studies (those with larger standard errors) compared to
meta-analysis. Moreover, since the expectation of the median
estimate~\eqref{eq:edgingtonpointest} is again a weighted average of the true
effects, it follows that Edgington's median estimate is unbiased when the true
effects coincide ($\theta_1 = \theta_2 = \theta$), while in case they differ,
the median estimate targets a weighted average of the true effects, though not
the same weighted average as targeted by meta-analysis.

The standard error of Edgington's median estimate is given by
\begin{equation}
  \label{eq:se2E}
  \sigma_{\text{E}} =
  \frac{\sqrt{2}}{1/\sigma_1 + 1/\sigma_2}
\end{equation}
and does not depend on the true effects, similar to meta-analysis but unlike the
two-trials rule. It is always larger than the meta-analytic standard
error~\eqref{eq:seMA}, see Appendix~\ref{app:ses}. Therefore, under effect
homogeneity, Edgington's method is less efficient than meta-analysis at
estimating the common effect. Under effect heterogeneity, however, the two
methods target different estimands, so a comparison of their standard errors may
not be meaningful. Finally, unlike meta-analysis, Edgington's standard error is
not always smaller or equal to either of the two trials' standard errors. This
is only the case if the standard error ratio is $\sqrt{2} - 1 \leq
\sigma_2/\sigma_1 \leq \sqrt{2} + 1$. For example, suppose $\sigma_1 = 0.5$ and
$\sigma_2 = 2$, then Edgington's standard error is $\sqrt{2}/(2 + 0.5) =
0.566$, which is greater than $\sigma_1$

In general, CIs for Edgington's method do not have closed-form solutions and
must be computed numerically. Nevertheless, as with Pearson's and Fisher's
methods, a closed-form combined estimation function is available when the effect
estimates and standard errors from both trials coincide ($\hat{\theta} =
\hat{\theta}_{1} = \hat{\theta}_{2}$ and $\sigma = \sigma_{1} = \sigma_{2}$),
which enables again analytical assessment of how the CI behaves in this
important scenario. In this case, the combined estimation function is
\begin{equation}
  \label{eq:edgingtonthesame}
  \hat{\mu}_{\text{E}}(a) =
  \begin{cases}
    \hat{\theta} + \sigma \, z_{\sqrt{a/2}} & \text{for} ~ a \leq 1/2
    \\ \hat{\theta} - \sigma \, z_{\sqrt{(1 - a)/2}} & \text{for} ~ a > 1/2 \\
  \end{cases}
\end{equation}
for the alternative "greater" and with the plus (minus) after $\hat{\theta}$
replaced with minus (plus) in~\eqref{eq:edgingtonthesame} for the alternative
"less". We can see that CIs obtained
from~\eqref{eq:edgingtonthesame} are symmetric and centered around the observed
effect estimate $\hat{\theta}$, similar to meta-analysis but unlike the
CIs from the two-trials rule, Tippett's, Fisher's, and Pearson's methods.
Yet, Edgington's CI is in this case narrower than the
meta-analytic CI. For example, Edgington's 95\% CI is
12.2\%
narrower than the corresponding meta-analytic 95\% CI. Panel A of
Figure~\ref{fig:extremeExamples} illustrates this as Edgington's CI is narrower
than the meta-analytic CI, although both are centered around the same effect
estimate. However, in case the trials' effect estimates are different,
Edgington's CI can also be much wider. For instance, in panel B of
Figure~\ref{fig:extremeExamples} where the trials produced very different
results, Edgington's CI is much wider than any of the other methods. This
suggests that Edgington's method reacts to heterogeneity by widening its CI to
include both trial effect estimates.

\subsubsection{Asymptotics}
Because the median estimate based on Edgington's method is a weighted average of
two normally distributed effect estimates, it is also normally distributed
\begin{align*}
  \hat{\mu}_{\text{E}}(1/2) \sim \mathrm{N}\left(\frac{\theta_1}{1 + \sqrt{c}} + \frac{\theta_2}{1 + 1/\sqrt{c}}, \, \frac{2}{(1/\sigma_1 + 1/\sigma_2)^2}\right)
\end{align*}
with relative variance $c = \sigma_1^2/\sigma_2^2$. As the median estimate has
its mean at the weighted average
\begin{align*}
\frac{\theta_1}{1 + \sqrt{c}} + \frac{\theta_2}{1 + 1/\sqrt{c}}
\end{align*}
it is clear that it will converge toward it as the standard errors decrease.
Whether the CI shrinks to this weighted average depends on whether the true
effects are equal. In case they are, it can be informally seen that the
CI~\eqref{eq:edgingtonthesame} will shrink to the common true effect, which is
illustrated in panel C of Figure~\ref{fig:extremeExamples}. However, when the
true effects differ, the CI will not shrink to a point but remain an interval
that always includes both true effects as the limiting combined estimation
function is
\begin{equation}
  \plim_{\sigma_1, \sigma_2 \downarrow 0} \hat{\mu}_{\text{E}}(a) =
  \begin{cases}
    \min\{\theta_{1}, \theta_{2}\} & \text{for} ~ a < 1/2 \\
    \dfrac{\theta_1}{1 + \sqrt{c}} + \dfrac{\theta_2}{1 + 1/\sqrt{c}} & \text{for} ~ a = 1/2 \\
    \max\{\theta_{1}, \theta_{2}\} & \text{for} ~ a > 1/2 \\
  \end{cases}
\end{equation}
see Appendix~\ref{app:approxmu} for details. This means that CIs based on
Edgington's method will asymptotically always include both true effects, even
when the trials' sample sizes become arbitrarily large, see panel D of
Figure~\ref{fig:extremeExamples} for an illustration. This behavior is
strikingly different from meta-analysis whose CI shrinks to a point at the
weighted average, even when the true effects are not the same.

\section{Applications}
\label{sec:application}
We will now illustrate combined \textit{p}-value functions, CIs, and median estimates
on data from two different pairs of clinical trials.

\subsection{The RESPIRE trials}
\begin{figure}[!htb]
\begin{knitrout}
\definecolor{shadecolor}{rgb}{0.969, 0.969, 0.969}\color{fgcolor}
\includegraphics[width=\maxwidth]{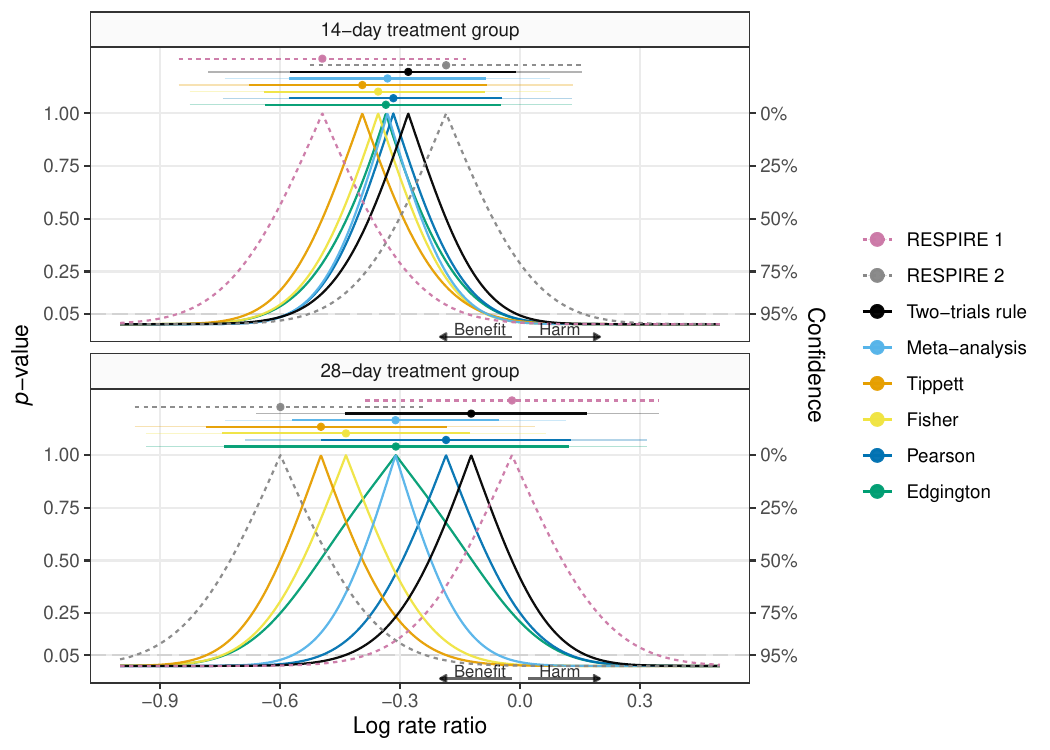} 
\end{knitrout}
\caption{Results of the RESPIRE trials \citep{Aksamit2018, DeSoyza2018,
    Chotirmall2018} for the effect of ciprofloxacin over 14 days (top) or 28
  days (bottom) compared to placebo for the treatment of non-cystic fibrosis
  bronchiectasis. The two-sided \textit{p}-value functions of the individual trials
  (dashed lines), and the combined \textit{p}-value functions (solid lines) based on
  the two-trials rule, fixed-effect meta-analysis, Tippett's, Fisher's,
  Pearson's, and Edgington's methods are shown along with corresponding median
  estimates and CIs (95\% and 99.875\% via
  telescope lines). All \textit{p}-values are based on the alternative
  "greater" and then converted to two-sided \textit{p}-values via the
  centrality function $2\min\{p, 1 - p\}$.}
\label{fig:twinstudies}
\end{figure}

We first revisit the RESPIRE trials \citep{Aksamit2018, DeSoyza2018,
  Chotirmall2018}, which were presented as motivating example in
Table~\ref{tab:respire} in the introduction. The trials investigated the effect
of ciprofloxacin in the treatment of non-cystic fibrosis bronchiectasis. Each
trial had two treatment groups (on/off treatment cycles of either 14 or 28 days
for 48 weeks) and two corresponding control groups. RESPIRE~1 showed a
substantial treatment effect in the 14-day treatment regimen (estimated log rate
ratio of $\log\widehat{\mathrm{RR}} = -0.49$ with 95\% CI
from $-0.85$ to $-0.13)$,
while the benefit was less pronounced in RESPIRE~2 ($\log\widehat{\mathrm{RR}} =
-0.18$ with 95\% CI from $-0.53$ to $0.16$). Surprisingly, this was reversed
for the 28-day regimens, with RESPIRE~2 showing a much stronger treatment effect
($\log\widehat{\mathrm{RR}} = -0.6$ with 95\% CI from
$-0.96$ to $-0.23)$ while
RESPIRE~1 showed almost no benefit ($\log\widehat{\mathrm{RR}} =
-0.02$ with 95\% CI from $-0.39$ to $0.35)$. Figure~\ref{fig:twinstudies}
shows the \textit{p}-value functions of the two studies (dashed lines) along with
different combined \textit{p}-value functions (solid lines) and corresponding point
estimates and CIs (top). Table~\ref{tab:tabrespirecombined} shows the results in
numerical form. Of note, all results were computed with our R package
\texttt{twotrials} and Appendix~\ref{app:rpackage} shows how the results for the
14-day treatment group can be reproduced.

\begin{table}[!htb]
  \centering
  \caption{Point estimates (with implicit weights), 95\% CIs (with widths), and
    \textit{p}-values for the RESPIRE trials \citep{Aksamit2018, DeSoyza2018,
      Chotirmall2018}. Note that weigths and CI widths are computed from
    unrounded numbers and may not always exactly correspond to the rounded point
    estimates and CIs.}
  \label{tab:tabrespirecombined}

\begin{tabular}{lccccc}
\toprule
 & \textbf{Log rate ratio} & \textbf{Weight RESPIRE 1} & \textbf{95\% CI} & \textbf{CI width} & \textbf{\textit{P}-value (one-sided)}\\
\midrule
\addlinespace[0.3em]
\multicolumn{6}{l}{\textit{14-day treatment group}}\\
\cellcolor[HTML]{F0F0F0}{\hspace{1em}RESPIRE 1} & \cellcolor[HTML]{F0F0F0}{$-0.49$} & \cellcolor[HTML]{F0F0F0}{$ $} & \cellcolor[HTML]{F0F0F0}{$-0.85$ to $-0.13$} & \cellcolor[HTML]{F0F0F0}{$\phantom{-}0.72$} & \cellcolor[HTML]{F0F0F0}{$\phantom{-}0.00351$}\\
\hspace{1em}RESPIRE 2 & $-0.18$ & $ $ & $-0.53$ to $\phantom{-}0.16$ & $\phantom{-}0.68$ & $\phantom{-}0.14400$\\
\midrule
\cellcolor[HTML]{F0F0F0}{\hspace{1em}Two-trials rule} & \cellcolor[HTML]{F0F0F0}{$-0.28$} & \cellcolor[HTML]{F0F0F0}{$\phantom{-}0.31$} & \cellcolor[HTML]{F0F0F0}{$-0.57$ to $-0.01$} & \cellcolor[HTML]{F0F0F0}{$\phantom{-}0.56$} & \cellcolor[HTML]{F0F0F0}{$\phantom{-}0.02073$}\\
\hspace{1em}Meta-analysis & $-0.33$ & $\phantom{-}0.47$ & $-0.58$ to $-0.08$ & $\phantom{-}0.49$ & $\phantom{-}0.00432$\\
\cellcolor[HTML]{F0F0F0}{\hspace{1em}Tippett} & \cellcolor[HTML]{F0F0F0}{$-0.39$} & \cellcolor[HTML]{F0F0F0}{$\phantom{-}0.68$} & \cellcolor[HTML]{F0F0F0}{$-0.68$ to $-0.08$} & \cellcolor[HTML]{F0F0F0}{$\phantom{-}0.59$} & \cellcolor[HTML]{F0F0F0}{$\phantom{-}0.00701$}\\
\hspace{1em}Fisher & $-0.35$ & $\phantom{-}0.55$ & $-0.64$ to $-0.09$ & $\phantom{-}0.55$ & $\phantom{-}0.00434$\\
\cellcolor[HTML]{F0F0F0}{\hspace{1em}Pearson} & \cellcolor[HTML]{F0F0F0}{$-0.32$} & \cellcolor[HTML]{F0F0F0}{$\phantom{-}0.43$} & \cellcolor[HTML]{F0F0F0}{$-0.58$ to $-0.04$} & \cellcolor[HTML]{F0F0F0}{$\phantom{-}0.53$} & \cellcolor[HTML]{F0F0F0}{$\phantom{-}0.01138$}\\
\hspace{1em}Edgington & $-0.34$ & $\phantom{-}0.49$ & $-0.64$ to $-0.05$ & $\phantom{-}0.59$ & $\phantom{-}0.01088$\\
\addlinespace[0.3em]
\multicolumn{6}{l}{\textit{28-day treatment group}}\\
\cellcolor[HTML]{F0F0F0}{\hspace{1em}RESPIRE 1} & \cellcolor[HTML]{F0F0F0}{$-0.02$} & \cellcolor[HTML]{F0F0F0}{$ $} & \cellcolor[HTML]{F0F0F0}{$-0.39$ to $\phantom{-}0.35$} & \cellcolor[HTML]{F0F0F0}{$\phantom{-}0.73$} & \cellcolor[HTML]{F0F0F0}{$\phantom{-}0.45699$}\\
\hspace{1em}RESPIRE 2 & $-0.60$ & $ $ & $-0.96$ to $-0.23$ & $\phantom{-}0.73$ & $\phantom{-}0.00064$\\
\midrule
\cellcolor[HTML]{F0F0F0}{\hspace{1em}Two-trials rule} & \cellcolor[HTML]{F0F0F0}{$-0.12$} & \cellcolor[HTML]{F0F0F0}{$\phantom{-}0.82$} & \cellcolor[HTML]{F0F0F0}{$-0.44$ to $\phantom{-}0.17$} & \cellcolor[HTML]{F0F0F0}{$\phantom{-}0.61$} & \cellcolor[HTML]{F0F0F0}{$\phantom{-}0.20884$}\\
\hspace{1em}Meta-analysis & $-0.31$ & $\phantom{-}0.50$ & $-0.57$ to $-0.05$ & $\phantom{-}0.52$ & $\phantom{-}0.00912$\\
\cellcolor[HTML]{F0F0F0}{\hspace{1em}Tippett} & \cellcolor[HTML]{F0F0F0}{$-0.50$} & \cellcolor[HTML]{F0F0F0}{$\phantom{-}0.18$} & \cellcolor[HTML]{F0F0F0}{$-0.79$ to $-0.18$} & \cellcolor[HTML]{F0F0F0}{$\phantom{-}0.60$} & \cellcolor[HTML]{F0F0F0}{$\phantom{-}0.00127$}\\
\hspace{1em}Fisher & $-0.44$ & $\phantom{-}0.28$ & $-0.75$ to $-0.12$ & $\phantom{-}0.62$ & $\phantom{-}0.00266$\\
\cellcolor[HTML]{F0F0F0}{\hspace{1em}Pearson} & \cellcolor[HTML]{F0F0F0}{$-0.18$} & \cellcolor[HTML]{F0F0F0}{$\phantom{-}0.72$} & \cellcolor[HTML]{F0F0F0}{$-0.50$ to $\phantom{-}0.13$} & \cellcolor[HTML]{F0F0F0}{$\phantom{-}0.62$} & \cellcolor[HTML]{F0F0F0}{$\phantom{-}0.12562$}\\
\hspace{1em}Edgington & $-0.31$ & $\phantom{-}0.50$ & $-0.74$ to $\phantom{-}0.12$ & $\phantom{-}0.86$ & $\phantom{-}0.10471$\\
\bottomrule
\end{tabular}

\end{table}

Looking at the combined point estimates, we can see that for both the 14-day and
28-day regimens, the estimate based on Tippett's method is the smallest (i.e.,
most anti-conservative for alternative = ``less''), while the estimate based on
the two-trials rule is the largest (i.e., the most conservative). A similar but
slightly attenuated pattern is seen for Fisher's (anti-conservative) and
Pearson's (conservative) methods, whereas the estimates from meta-analysis and
Edgington's method are almost identical and fall somewhere between the
individual trials' effect estimates. All point estimates are thus consistent
with the theoretically expected behavior of the methods.

It is interesting to consider the median estimate as a weighted average of the
trial specific point estimates, and to determine the corresponding (implicit)
weights. Table~\ref{tab:tabrespirecombined} reports the weight of the point
estimate from RESPIRE 1 toward the median estimate (the weight from RESPIRE 2 is
one minus the weight from RESPIRE 1).
The more extreme estimate (RESPIRE 1 in the 14-day group, and RESPIRE 2 in the
28-day group) contributes more to Tippett's and Fisher's estimates and less to
Pearson's and the two-trials rule estimates, which aligns with the expected
behavior. Similarly, the weight of RESPIRE 1 is slightly larger for Edgington's
method than meta-analysis because Edgington's estimate gives more weights to
trials with larger standard errors due to its inverse standard error weighting.

Looking at the CIs, we can see that meta-analysis produces narrower CIs than the
other methods for both treatment regimens. The widest CIs are produced by
Edgington's method. For the 28-day regimen, Edgington's 95\% CI is the only method
that includes both trial effect estimates, and as a result is even wider than
the CIs from the individual trials, reflecting the apparent heterogeneity.
Looking at the decision based on the CIs, we can see that for the 14-day
regimens, all 95\% CIs exclude a log rate ratio of zero, the value of no effect,
while all 99.875\% CIs include it. However, for
the 28-day regimens, the 95\% CIs from meta-analysis, Fisher's, and Tippett's
methods exclude zero. The other method's 95\% CIs include zero, but only
Edgington's method includes also the point estimate from RESPIRE~2. Finally, the
99.875\% CIs of all methods include zero, thus
leading to identical decisions at the one-sided $0.025^2 = 0.000625$
level. Note that for each method, the decision based on the CI is compatible
with the combined \textit{p}-values in Table~\ref{tab:tabrespirecombined}, for example,
a 99.875\% CI excludes a log rate ratio of zero
only if also the combined one-sided \textit{p}-value is less than $0.000625$.

\subsection{The ORBIT trials}
\begin{figure}[!tb]
\begin{knitrout}
\definecolor{shadecolor}{rgb}{0.969, 0.969, 0.969}\color{fgcolor}
\includegraphics[width=\maxwidth]{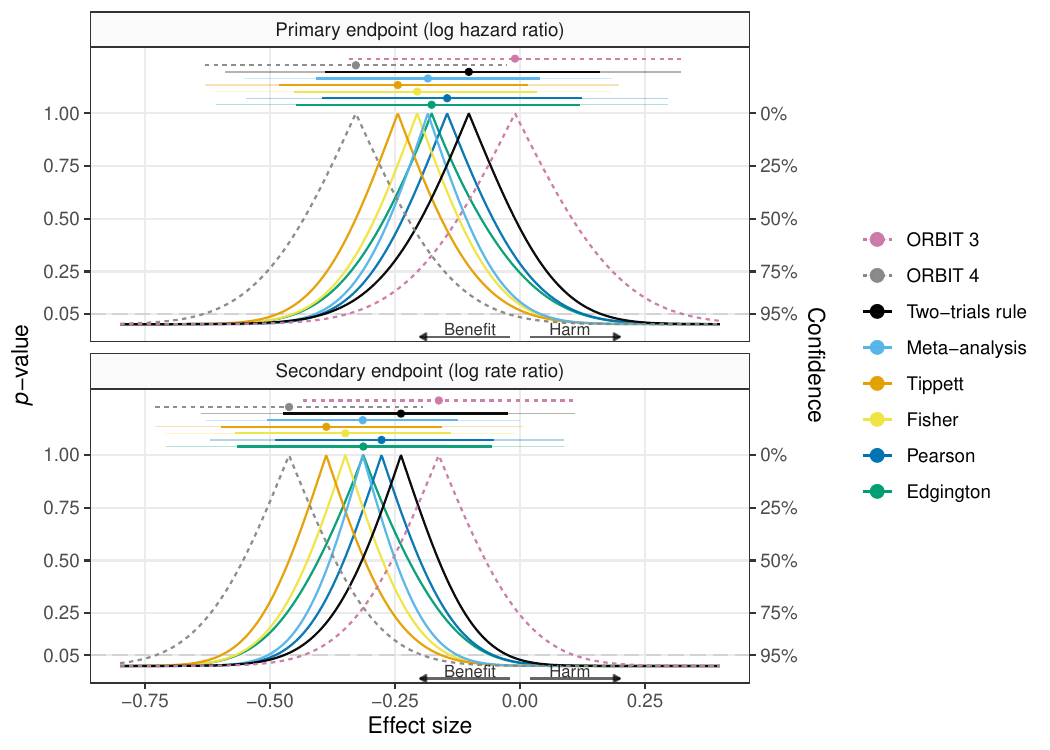} 
\end{knitrout}
\caption{Results of the ORBIT trials \citep{Haworth2019} for the effect of
  ciprofloxacin in patients with non-cystic fibrosis bronchiectasis and chronic
  lung infection with \textit{Pseudomonas aeruginosa}. The two-sided \textit{p}-value
  functions of the individual trials (dashed lines), and the combined \textit{p}-value
  functions (solid lines) based on the two-trials rule, fixed-effect
  meta-analysis, Tippett's, Fisher's, Pearson's, and Edgington's methods are
  shown along with corresponding median estimates and CIs (95\% and
  99.875\% via telescope lines). All \textit{p}-values
  are based on the alternative "greater" and then converted to two-sided
  \textit{p}-values via the centrality function $2\min\{p, 1 - p\}$.}
\label{fig:ORBIT}
\end{figure}

Another pair of clinical trials that investigated the effect of ciprofloxacin
are the ORBIT~3 and ORBIT~4 trials \citep{Haworth2019}. The trials assessed the
effect of inhaled liposomal ciprofloxacin compared to placebo in patients with
non-cystic fibrosis bronchiectasis and chronic lung infection with
\textit{Pseudomonas aeruginosa}. Like the RESPIRE trials, the ORBIT trials also
showed considerable heterogeneity. Figure~\ref{fig:ORBIT} shows \textit{p}-values,
point estimates, and CIs for the primary endpoint (time to the first
exacerbation; effect quantified with a log hazard ratio) and a secondary
endpoint (frequency of exacerbations; effect quantified with a log rate ratio).
Table~\ref{tab:taborbitcombined} gives numerical summaries.

We see that there is substantial heterogeneity for the primary endpoint, with
the point estimate from ORBIT 3 close to zero ($\log\widehat{\mathrm{HR}} =
-0.01$ with 95\% CI from $-0.34$ to
$0.32$), whereas the estimate from ORBIT 4 indicates a
more beneficial treatment effect ($\log\widehat{\mathrm{HR}} =
-0.33$ with 95\% CI from $-0.63$ to
$-0.03$). While the theoretically expected patterns of
the different median estimates and CIs are visible, the qualitative decisions
based on all the different combination methods are the same at both the $0.025$
and $0.025^2$ levels.

\begin{table}[!htb]
  \centering
  \caption{Point estimates (with implicit weights), 95\% CIs (with widths), and
    \textit{p}-values for the ORBIT trials \citep{Haworth2019}. Note that
    weigths and CI widths are computed from unrounded numbers and may not always
    exactly correspond to the rounded point estimates and CIs.}
  \label{tab:taborbitcombined}

\begin{tabular}{lccccc}
\toprule
 & \textbf{Log rate ratio} & \textbf{Weight ORBIT 3} & \textbf{95\% CI} & \textbf{CI width} & \textbf{\textit{P}-value (one-sided)}\\
\midrule
\addlinespace[0.3em]
\multicolumn{6}{l}{\textit{\textit{Primary endpoint (log hazard ratio)}}}\\
\cellcolor[HTML]{F0F0F0}{\hspace{1em}ORBIT 3} & \cellcolor[HTML]{F0F0F0}{$-0.01$} & \cellcolor[HTML]{F0F0F0}{$ $} & \cellcolor[HTML]{F0F0F0}{$-0.34$ to $\phantom{-}0.32$} & \cellcolor[HTML]{F0F0F0}{$\phantom{-}0.66$} & \cellcolor[HTML]{F0F0F0}{$\phantom{-}0.47636$}\\
\hspace{1em}ORBIT 4 & $-0.33$ & $ $ & $-0.63$ to $-0.03$ & $\phantom{-}0.60$ & $\phantom{-}0.01657$\\
\midrule
\cellcolor[HTML]{F0F0F0}{\hspace{1em}Two-trials rule} & \cellcolor[HTML]{F0F0F0}{$-0.10$} & \cellcolor[HTML]{F0F0F0}{$\phantom{-}0.71$} & \cellcolor[HTML]{F0F0F0}{$-0.39$ to $\phantom{-}0.16$} & \cellcolor[HTML]{F0F0F0}{$\phantom{-}0.55$} & \cellcolor[HTML]{F0F0F0}{$\phantom{-}0.22692$}\\
\hspace{1em}Meta-analysis & $-0.18$ & $\phantom{-}0.45$ & $-0.41$ to $\phantom{-}0.04$ & $\phantom{-}0.45$ & $\phantom{-}0.05305$\\
\cellcolor[HTML]{F0F0F0}{\hspace{1em}Tippett} & \cellcolor[HTML]{F0F0F0}{$-0.24$} & \cellcolor[HTML]{F0F0F0}{$\phantom{-}0.26$} & \cellcolor[HTML]{F0F0F0}{$-0.48$ to $\phantom{-}0.02$} & \cellcolor[HTML]{F0F0F0}{$\phantom{-}0.50$} & \cellcolor[HTML]{F0F0F0}{$\phantom{-}0.03286$}\\
\hspace{1em}Fisher & $-0.21$ & $\phantom{-}0.39$ & $-0.45$ to $\phantom{-}0.03$ & $\phantom{-}0.49$ & $\phantom{-}0.04610$\\
\cellcolor[HTML]{F0F0F0}{\hspace{1em}Pearson} & \cellcolor[HTML]{F0F0F0}{$-0.15$} & \cellcolor[HTML]{F0F0F0}{$\phantom{-}0.57$} & \cellcolor[HTML]{F0F0F0}{$-0.40$ to $\phantom{-}0.12$} & \cellcolor[HTML]{F0F0F0}{$\phantom{-}0.52$} & \cellcolor[HTML]{F0F0F0}{$\phantom{-}0.14328$}\\
\hspace{1em}Edgington & $-0.18$ & $\phantom{-}0.48$ & $-0.45$ to $\phantom{-}0.12$ & $\phantom{-}0.57$ & $\phantom{-}0.12149$\\
\addlinespace[0.3em]
\multicolumn{6}{l}{\textit{\textit{Secondary endpoint (log rate ratio)}}}\\
\cellcolor[HTML]{F0F0F0}{\hspace{1em}ORBIT 3} & \cellcolor[HTML]{F0F0F0}{$-0.16$} & \cellcolor[HTML]{F0F0F0}{$ $} & \cellcolor[HTML]{F0F0F0}{$-0.43$ to $\phantom{-}0.11$} & \cellcolor[HTML]{F0F0F0}{$\phantom{-}0.54$} & \cellcolor[HTML]{F0F0F0}{$\phantom{-}0.12083$}\\
\hspace{1em}ORBIT 4 & $-0.46$ & $ $ & $-0.73$ to $-0.19$ & $\phantom{-}0.54$ & $\phantom{-}0.00036$\\
\midrule
\cellcolor[HTML]{F0F0F0}{\hspace{1em}Two-trials rule} & \cellcolor[HTML]{F0F0F0}{$-0.24$} & \cellcolor[HTML]{F0F0F0}{$\phantom{-}0.75$} & \cellcolor[HTML]{F0F0F0}{$-0.47$ to $-0.02$} & \cellcolor[HTML]{F0F0F0}{$\phantom{-}0.45$} & \cellcolor[HTML]{F0F0F0}{$\phantom{-}0.01460$}\\
\hspace{1em}Meta-analysis & $-0.31$ & $\phantom{-}0.49$ & $-0.51$ to $-0.12$ & $\phantom{-}0.38$ & $\phantom{-}0.00062$\\
\cellcolor[HTML]{F0F0F0}{\hspace{1em}Tippett} & \cellcolor[HTML]{F0F0F0}{$-0.39$} & \cellcolor[HTML]{F0F0F0}{$\phantom{-}0.25$} & \cellcolor[HTML]{F0F0F0}{$-0.60$ to $-0.16$} & \cellcolor[HTML]{F0F0F0}{$\phantom{-}0.44$} & \cellcolor[HTML]{F0F0F0}{$\phantom{-}0.00072$}\\
\hspace{1em}Fisher & $-0.35$ & $\phantom{-}0.38$ & $-0.57$ to $-0.14$ & $\phantom{-}0.43$ & $\phantom{-}0.00048$\\
\cellcolor[HTML]{F0F0F0}{\hspace{1em}Pearson} & \cellcolor[HTML]{F0F0F0}{$-0.28$} & \cellcolor[HTML]{F0F0F0}{$\phantom{-}0.62$} & \cellcolor[HTML]{F0F0F0}{$-0.49$ to $-0.05$} & \cellcolor[HTML]{F0F0F0}{$\phantom{-}0.44$} & \cellcolor[HTML]{F0F0F0}{$\phantom{-}0.00765$}\\
\hspace{1em}Edgington & $-0.31$ & $\phantom{-}0.50$ & $-0.57$ to $-0.06$ & $\phantom{-}0.51$ & $\phantom{-}0.00734$\\
\bottomrule
\end{tabular}

\end{table}

Looking at the secondary endpoint, there is also considerable heterogeneity
between the results from ORBIT 3 ($\log\widehat{\mathrm{RR}} =
-0.16$ with 95\% CI from $-0.43$ to
$0.11$) and ORBIT 4 ($\log\widehat{\mathrm{RR}} =
-0.46$ with 95\% CI from $-0.73$ to
$-0.19$) leading to some more
noticeable qualitative differences between the methods. That is, the
99.875\% CIs from meta-analysis and Fisher's
method exclude a log rate ratio of zero while the remaining methods include it,
leading to different decisions at the $0.025^2$ level. Again, Edgington's CI is
much wider than the others due to the substantial heterogeneity.

In summary, the analyses of the RESPIRE and ORBIT trials showed how combined
\textit{p}-value functions allow us to obtain point estimates, CIs, and
\textit{p}-values that are inherently compatible. They also showed that
different combination methods can lead to different inferences and decisions,
especially in the presence of between-trial heterogeneity, highlighting the need
to think about the estimand of interest.

\section{Extension to more than two trials}
\label{sec:extensions}
The methods discussed so far have focused on the setting where only two trials
are available, but in practice it may happen that investigators want to assess
the combined evidence from more than two trials. In this context, Rosenkranz
\citep{Rosenkranz2023} suggested that decision rules should maintain the type I
error rate of the two-trials rule for two studies $\alpha^2$, even if there are
more than two studies. This can be implemented using combined \textit{p}-value
functions, as all methods considered before can be generalized to more than two
trials \citep{XieSingh2013, Held_etal2024b}. A decision rule can then be based
on the combined one-sided \textit{p}-value for the null hypothesis of no effect
or a $(1 - 2\alpha^2)\times 100\%$ CI obtained from a combined \textit{p}-value
function. In addition, a point estimate and 95\% CI can be used to summarize the
combined evidence.

While all point estimates and CIs in this setting can be computed numerically,
some of the analytical results derived earlier generalize to more than two
studies. Specifically, closed-form median estimates and CIs remain available for
the two-trials rule, Tippett's method, and meta-analysis, whereas such
closed-form solutions are not available for Fisher's, Pearson's, and Edgington's
methods \citep{Held_etal2024b}. In particular, for Edgington's method, one might
expect the median estimate~\eqref{eq:edgingtonpointest} to generalize by
incorporating additional effect estimates with inverse standard error weights.
However, a comparison with numerically computed median estimates showed that
this is not the case. Thus, the inverse standard error weighted average
in~\eqref{eq:edgingtonpointest} corresponds to Edgington's median estimate only
in the setting of two trials.

\begin{figure}[!htb]
\begin{knitrout}
\definecolor{shadecolor}{rgb}{0.969, 0.969, 0.969}\color{fgcolor}
\includegraphics[width=\maxwidth]{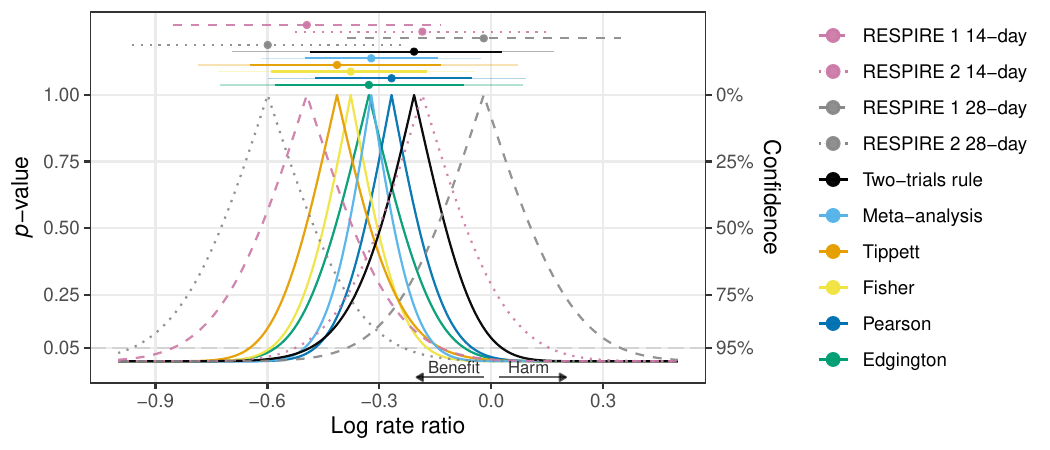} 
\end{knitrout}
\caption{Combined \textit{p}-value function based on all four estimates from the
  RESPIRE trials \citep{Aksamit2018, DeSoyza2018, Chotirmall2018} along with
  corresponding median estimates and CIs (95\% and 99.875\% via telescope
  lines).}
\label{fig:respireall}
\end{figure}

Figure~\ref{fig:respireall} shows \textit{p}-value functions that combine all
four results from the two RESPIRE trials, as also done by Chotirmall and
Chalmers \citep{Chotirmall2018} with fixed-effect meta-analysis. Looking at the
median estimates, we see the same patterns as when the methods are applied to
only two trials; The median estimates from meta-analysis and Edgington's method
are somewhere in between the trials' individual estimates. The median estimates
based on Fisher's and Tippett's method are the most anti-conservative, and the
median estimates based on Pearson's method and the two-trials rule are the most
conservative. A decision rule that maintains the $\alpha^2 = 0.025^2$ type I
error rate of the two trials rule could now be defined by flagging drug efficacy
when the $(1 - 2\alpha^2) \times 100\% = 99.875\%$ excludes the null value.
Following this rule, we can see that Fisher's method and meta-analysis flag
efficacy while the remaining methods do not. In sum, this example illustrates
how combined \textit{p}-value functions can be applied to more than two trials,
while allowing to maintain the same type I error rate as for two trials.

\section{Discussion}
\label{sec:discussion}
The two-trials rule has been widely discussed in the literature but discussions
have mostly focused on hypothesis testing characteristics, such as power or type
I error rate \citep{Fisher1999, Fisher1999b, Lu2001, Rosenkranz2002, Maca2002,
  Shao2002, Shun2005, vanRavenzwaaij2017, kennedySchaffer:2017, Senn2021,
  Zhan2022, Rosenkranz2023, Held2024}. In this paper, we took a different
perspective, systematically examining the two-trials rule and alternative
methods in terms of effect estimation. By casting them in a combined \textit{p}-value
function framework, we derived compatible \textit{p}-values, confidence intervals, and
point estimates. These quantities are compatible in the sense that the two-sided
\textit{p}-value for a null value is less than $\alpha$ if and only if the null value
is excluded by the $(1 - \alpha) \times 100\%$ CI, and that the point estimate
is contained in the CI at any confidence level. While meta-analytic effect
estimates, CIs, and \textit{p}-values have been well studied, our novel results enable
computation of CIs and effect estimates based on the two-trials rule.
Investigators could therefore report not only individual trial \textit{p}-values
(essentially the two-trials rule) and meta-analytic estimates but also point
estimates and CIs based on the two-trials rule.

Our findings also clarify how different \textit{p}-value combination methods
implicitly target different estimands. Reassuringly, under effect homogeneity
(i.e., the same true effect in both trials), all methods yield consistent point
estimates and CIs that shrink toward the true effect as standard errors
decrease, although some show bias. Theoretically, meta-analysis has the smallest
variance among all unbiased estimators (attaining the Cramér-Rao lower bound)
and may be preferred. However, under effect heterogeneity -- arguably the more
realistic scenario -- it is less clear which method should be recommended. The
two-trials rule and Pearson's method are conservative (targeting the less
extreme effect), Fisher's and Tippett's methods are anti-conservative (targeting
the more extreme effect), while Edgington's method and meta-analysis are
balanced (targeting a weighted average). This raises an important question: What
kind of effect is of scientific interest when the true trial effects differ? If
the investigators are interested in the less extreme effect -- arguably a
sensible choice when the effects relate to a medical treatment with potential
side effects -- then the two-trials rule and Pearson's method seem reasonable.
On the other hand, a weighted average effect, as targeted by meta-analysis and
Edgington's method, might be a relevant estimand if it is representative for a
larger population \citep{Rice_etal2018}. Finally, the more extreme effect might
be the relevant estimand if the maximum achievable benefit of a treatment is of
scientific interest, in which case Tippett's and Fisher's methods might be
reasonable choices. 
This parallels the findings of Heard and Rubin-Delanchy \citep{Heard2018}, who
showed that many \textit{p}-value combination methods are equivalent to a
likelihood ratio test for specific alternative hypotheses. This means that each
such method can be most powerful under certain conditions. Therefore,
researchers must carefully reflect which alternative hypothesis is most relevant
to their application -- just as they need to reflect on choosing an appropriate
estimand -- to select a suitable combination method.

Beyond theoretical considerations, practical issues must be addressed. A major
concern is that if the effect estimates from both trials are the same, the
two-trials rule, Tippett's, Fisher's, and Pearson's methods all produce
counterintuitive effect estimates that differ from the one observed in both
trials. Such point estimates are unintuitive and difficult to communicate to
non-statisticians. Moreover, only Edgington's method and meta-analysis produce
the same combined estimate and two-sided confidence interval in case the
alternative of the combined $p$-values is changed \citep{Held_etal2024b}, which
seems another practically desired property. From this perspective, Edgington's
method and meta-analysis may be preferable. In particular, Edgington's method
can also account for effect heterogeneity by widening its CI when there is
heterogeneity and asymptotically always includes both effects. However, this is
traded off with a less efficient median estimate under effect homogeneity, whose
standard error can even be larger than those from both trials if they greatly
differ. Finally, another practical challenge is aligning decisions based on a
one-sided combined \textit{p}-value thresholded at $\alpha^2$ with two-sided
CIs. This requires using a $(1 - 2\alpha^{2})\times 100\%$ confidence level.
However, in many fields, researchers are not used to such confidence levels, so
we suggest to report both a more conventional level (e.g., 95\%) along with $(1
- 2\alpha^{2}) \times 100\%$ via telescope-style CIs, as well as the underlying
\textit{p}-value function, as in
Figures~\ref{fig:twinstudies}--\ref{fig:respireall}. The idea of telescope-style
CIs is not new but has been suggested before in different contexts
\citep{Louis2008}.

A broader issue is the question of whether two trials are actually necessary. If
the designs of the two trials are so similar that they can be considered
exchangeable (``direct replications'' \citep{Nosek2017b}), there are various
arguments in favor of conducting one large trial instead of two smaller ones
\citep{Fisher1999,Senn2021}. Also our study demonstrates that having two trials
instead of one makes estimation more complicated. Conversly, if the trial
designs differ significantly (``conceptual replications'' \citep{Nosek2017b},
e.g., if they use different endpoints or populations), achieving success in both
trials may provide more robust evidence of treatment efficacy. From this
perspective, it is sensible to design the trials differently to some extent
\citep{Senn2021}. However, there is perhaps a limit to how different the trials
can be, as when there is too much heterogeneity, combining the effect estimates
into a single number would no longer be meaningful.

Our results have broader implications beyond the two-trial setting. Methods for
combining \textit{p}-values are also used in adaptive trials, where they enable
combination of stage-wise \textit{p}-values \citep{Bauer1994, Lehmacher1999,
  Bauer1999}. Combined \textit{p}-value functions can be generalized to more
than two studies, making them applicable to meta-analysis \citep{XieSingh2013,
  Held_etal2024b}. They can also be applied to replication and real-world
evidence studies, where the two-trials rule (under different names such as
significance criterion or vote-counting) is used to assess the replicability of
original findings \citep{Bartlett2019, Wang2022, Held_etal2024}. In all these
scenarios, we may consider combined \textit{p}-value functions for parameter
estimation, but in each application researchers must also decide which
combination method has the statistical properties to estimate the scientific
effect of interest. Future research may also examine other combination methods
beyond the ones considered here, such as the inverse chi-square method
\citep{Lancaster1961, HedgesOlkin1985}, the harmonic mean $\chi^2$ test
\citep{Held2020b}, the Cauchy combination test \citep{Liu2019}, random-effects
meta-analysis \citep{Borenstein2010}, and combining \textit{p}-value functions
that are based on the exact distribution of the data rather than normality,
e.g., the \textit{p}-value function based on Fisher's exact test with
mid-\textit{p} correction \citep{Schweder2013, Held_etal2024b}. Additionally,
fixed-effect meta-analysis has a Bayesian interpretation, corresponding to
posterior inferences assuming equal true study effects and a flat prior
distribution. Investigating whether other \textit{p}-value combination methods
have similar Bayesian justifications could be an interesting avenue for future
work. To sum up, combined \textit{p}-value functions provide a unified approach
for combining results from two trials that can be further developed
theoretically. Moreover, our software implementation allows researchers to
conveniently apply these methods in practice.

\bmsection*{Acknowledgments}

We thank the editor, associate editor, Stephen Senn, and another anonymous
reviewer for valuable comments that led to numerous additions and improvements.
The acknowledgment of these individuals does not imply their endorsement of the
paper.

\bmsection*{Conflict of interest}
We declare no conflict of interest.

\bmsection*{Software and data} Data from the RESPIRE trials were extracted from
Table~3 in De Soyza et al. \citep{DeSoyza2018} and Table~3 in Aksamit et al.
\citep{Aksamit2018}. Data from the ORBIT trials were extracted from page~219 in
Haworth et al. \citep{Haworth2019}. Code and data to reproduce all numbers,
tables, and figures are openly available at
\url{https://github.com/SamCH93/twotrials}. Spreadsheets containing the numbers
from Tables~\ref{tab:tabrespirecombined} and~\ref{tab:taborbitcombined} with
higher precision are also available at the repository. A snapshot of the
repository at the time of writing is available at
\url{https://doi.org/10.5281/zenodo.15017483}. The R package \texttt{twotrials}
implementing combined \textit{p}-value function inference for two trials is
available at \url{https://doi.org/10.32614/CRAN.package.twotrials}. We used the
statistical programming language R version 4.4.1 (2024-06-14) for
analyses \citep{R} along with the \texttt{confMeta} \citep{confMeta2024},
\texttt{ggplot2} \citep{Wickham2016}, \texttt{kableExtra} \citep{Zhu2024},
\texttt{dplyr} \citep{Wickham2023}, and \texttt{knitr} \citep{Xie2015} packages.

\bibliography{bibliography.bib}

\appendix

\bmsection{The R package twotrials}
\label{app:rpackage}
We have developed the \texttt{twotrials} R package for easily conducting
combined \textit{p}-value function inference based on the parameter estimates
(with standard errors) from two trials. The package can be installed from the
Comprehensive R Archive Network (CRAN) via the R command
\texttt{install.packages("twotrials")}.

For every \textit{p}-value combination method discussed in this paper, the package
provides a combined \textit{p}-value function (e.g., \texttt{pEdgington}) and a
combined estimation function (e.g., \texttt{muEdgington}). While these can be
used to manually compute \textit{p}-values and parameter estimates, the convenience
function \texttt{twotrials} automatically computes estimates and \textit{p}-values
based on all methods, and allows for easy printing and plotting of the results.
The following code chunk illustrates its usage by reproducing the results for
the 14-day treatment group from Table~\ref{tab:tabrespirecombined}.

\begin{knitrout}\scriptsize
\definecolor{shadecolor}{rgb}{0.969, 0.969, 0.969}\color{fgcolor}\begin{kframe}
\begin{alltt}
\hlkwd{library}\hldef{(twotrials)} \hlcom{# load package}

\hlcom{## combine logRR estimates from RESPIRE trials}
\hldef{results} \hlkwb{<-} \hlkwd{twotrials}\hldef{(}\hlkwc{null} \hldef{=} \hlnum{0}\hldef{,} \hlkwc{t1} \hldef{=} \hlopt{-}\hlnum{0.4942}\hldef{,} \hlkwc{t2} \hldef{=} \hlopt{-}\hlnum{0.1847}\hldef{,} \hlkwc{se1} \hldef{=} \hlnum{0.1833}\hldef{,}
                     \hlkwc{se2} \hldef{=} \hlnum{0.1738}\hldef{,} \hlkwc{alternative} \hldef{=} \hlsng{"less"}\hldef{,} \hlkwc{level} \hldef{=} \hlnum{0.95}\hldef{)}
\hlkwd{print}\hldef{(results,} \hlkwc{digits} \hldef{=} \hlnum{2}\hldef{)} \hlcom{# print summary of results}
\end{alltt}
\begin{verbatim}
## INDIVIDUAL RESULTS
##    Trial Lower CL Estimate Upper CL P-value
##  Trial 1    -0.85    -0.49    -0.13  0.0035
##  Trial 2    -0.53    -0.18     0.16  0.1440
## 
## COMBINED RESULTS
##           Method Lower CL Estimate Upper CL P-value   W1   W2
##  Two-trials rule    -0.57    -0.28   -0.011  0.0207 0.31 0.69
##    Meta-analysis    -0.58    -0.33   -0.084  0.0043 0.47 0.53
##          Tippett    -0.68    -0.39   -0.084  0.0070 0.68 0.32
##           Fisher    -0.64    -0.35   -0.087  0.0043 0.55 0.45
##          Pearson    -0.58    -0.32   -0.044  0.0114 0.43 0.57
##        Edgington    -0.64    -0.34   -0.048  0.0109 0.49 0.51
## 
## NOTES 
## Confidence level: 95%
## Null value: 0
## Alternative: less
\end{verbatim}
\end{kframe}
\end{knitrout}

Note that, for each combined estimate, the function also returns the weights
$w_1$ (\texttt{W1}) and $w_2$ (\texttt{W2}). These represent the implicit linear
weights of the point estimates from trials 1 and 2 towards the combined
estimate, i.e., $\hat{\mu}(1/2) = w_1 \hat{\theta}_1 + w_2 \hat{\theta}_2$. Such
weights aid interpretation by indicating how close each trial estimate is to the
combined estimate. Finally, applying the plot function to the resulting object
makes it easy to display the combined \textit{p}-value functions, as
demonstrated below.

\begin{knitrout}\scriptsize
\definecolor{shadecolor}{rgb}{0.969, 0.969, 0.969}\color{fgcolor}\begin{kframe}
\begin{alltt}
\hlkwd{plot}\hldef{(results,} \hlkwc{xlim} \hldef{=} \hlkwd{c}\hldef{(}\hlopt{-}\hlnum{1}\hldef{,} \hlnum{0.5}\hldef{),} \hlkwc{two.sided} \hldef{=} \hlnum{TRUE}\hldef{)} \hlcom{# plot p-value functions}
\end{alltt}
\end{kframe}
\includegraphics[width=\maxwidth]{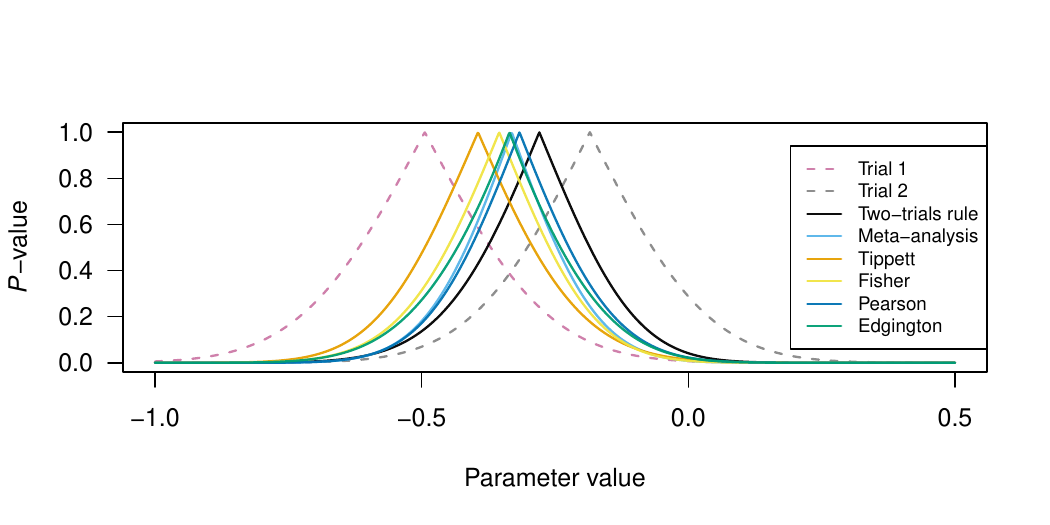} 
\end{knitrout}

\bmsection{Technical details}
\label{app:technicaldetails}
This appendix contains technical details on the derivation of results from the
main paper.

\bmsubsection{Expectation of the combined estimation function}
\label{app:expectation}
Consider the random variables
\begin{align}
  \label{eq:XandY}
  &X = \min\{\hat{\theta}_1 + \sigma_1 \, q, \hat{\theta}_2 + \sigma_2 \, q\}&
   &\text{and}&
  &Y = \max\{\hat{\theta}_1 - \sigma_1 \, q, \hat{\theta}_2 - \sigma_2 \, q\}.&
\end{align}
Table~\ref{tab:constant} shows that for certain choices of the constant $q$, $X$
and $Y$ are equal to the combined estimation functions from the two-trials
rule~\eqref{eq:2trialsest} and Tippett's method~\eqref{eq:tippest} in the main
paper, and the approximate combined estimation functions from
Fisher's~\eqref{eq:fisherest2}, Pearson's~\eqref{eq:pearsonest2}, and
Edgington's methods~\eqref{eq:edgingtonprecise} and~\eqref{eq:edgingtonprecise2}
discussed below. Note that for Edgington's method (with $a = 1/2$) and
meta-analysis, the distribution of the combined estimation function is normal
and its expectation is therefore known and need not be derived here.

\begin{table}[!htb]
  \centering

  \caption{Constants $q$ for which $X$ or $Y$ are equal to the combined
    estimation function of a specific method.}
  \label{tab:constant} \rowcolors{1}{}{gray!15}

  \begingroup
  \renewcommand{\arraystretch}{1.3}
  \begin{tabular}{l l l}
    \toprule
    \multicolumn{1}{c}{\textbf{Method}} & \multicolumn{1}{c}{\textbf{Alternative ``greater''}} & \multicolumn{1}{c}{\textbf{Alternative ``less''}} \\

    \midrule
    Two-trials rule & $X$ with $q = z_{\sqrt{a}}$ & $Y$ with $q = z_{\sqrt{a}}$\\

    Tippett & $Y$ with $q = z_{\sqrt{1 - a}}$ & $X$ with $q = z_{\sqrt{1 - a}}$\\

    Fisher (approximate) & $Y$ with $q = -z_{\exp\{-\chi^2_4(1 - a)/2\}}$ & $X$ with $q = -z_{\exp\{-\chi^2_4(1 - a)/2\}}$ \\

    Pearson (approximate) & $X$ with $q = -z_{\exp\{-\chi^2_4(a)/2\}}$ & $Y$ with $q = -z_{\exp\{-\chi^2_4(a)/2\}}$ \\

    Edgington (approximate, $a < 1/2$) & $X$ with $q = z_{\sqrt{2a}}$ & $Y$ with $q = z_{\sqrt{2a}}$ \\

    Edgington (approximate, $a > 1/2$) & $Y$ with $q = z_{\sqrt{2(1 - a)}}$ & $X$ with $q = z_{\sqrt{2(1 - a)}}$ \\
    \bottomrule
  \end{tabular}
  \endgroup
\end{table}

According to the assumptions stated at the beginning of
Section~\ref{sec:combinedp}, $\hat{\theta}_1$ and $\hat{\theta}_2$ are
independent normal random variables with means $\theta_1, \theta_2$ and
variances $\sigma_1^2,\sigma_2^2$. We can therefore use the results from
Nadarajah and Kotz \citep{Nadarajah2008} on closed-form expressions for the
moments of minima and maxima of bivariate Gaussian random vectors. That is,
using their equations (9) and (11), it can be shown that the expectations of $X$
and $Y$ are given by
\begin{align*}
  E(X) =&
  (\theta_1 + \sigma_1 \, q) \times \Phi\left(\frac{\theta_2 - \theta_1 + q (\sigma_2 - \sigma_1)}{\sqrt{\sigma_1^2 + \sigma_2^2}}\right)
  + (\theta_2 + \sigma_2 \, q) \times \Phi\left(\frac{\theta_1 - \theta_2 + q (\sigma_1 - \sigma_2)}{\sqrt{\sigma_1^2 + \sigma_2^2}}\right)  \\
  &- \sqrt{\sigma_1^2 + \sigma_2^2} \times \phi\left(\frac{\theta_2 - \theta_1 + q (\sigma_2 - \sigma_1)}{\sqrt{\sigma_1^2 + \sigma_2^2}}\right)
\end{align*}
and
\begin{align*}
  E(Y) =&
  (\theta_1 - \sigma_1 \, q) \times \Phi\left(\frac{\theta_1 - \theta_2 + q (\sigma_2 - \sigma_1)}{\sqrt{\sigma_1^2 + \sigma_2^2}}\right)
  + (\theta_2 - \sigma_2 \, q) \times  \Phi\left(\frac{\theta_2 - \theta_1 + q (\sigma_1 - \sigma_2)}{\sqrt{\sigma_1^2 + \sigma_2^2}}\right) \\
  &+ \sqrt{\sigma_1^2 + \sigma_2^2} \times \phi\left(\frac{\theta_1 - \theta_2 + q (\sigma_2 - \sigma_1)}{\sqrt{\sigma_1^2 + \sigma_2^2}}\right),
\end{align*}
respectively. The expectation of the combined estimation function from a
specific method are thus obtained by setting the constant $q$ to the
corresponding value. For example, the expectation of the median estimate ($a =
1/2$) from the two-trials rule with alternative ``greater'' and $\sigma_1 =
\sigma_2 = \sigma$ in equation~\eqref{eq:exp2tr} is obtained from $E(X)$ with $q
= z_{\sqrt{1/2}}$.

\bmsubsection{Median estimate standard errors}
\label{app:ses}
Since the median estimates from meta-analysis and Edgington's method are simple
linear combinations of the trial effect estimates, their standard errors can be
straightforwardly derived to be
\begin{align*}
  &\sigma_{\text{MA}} 
  = \frac{1}{\sqrt{1/\sigma^2_1 + 1/\sigma^2_2}}& &\text{and}& &\sigma_{\text{E}} 
  = \frac{\sqrt{2}}{1/\sigma_1 + 1/\sigma_2}.&
\end{align*}
By applying algebraic manipulations to $\sigma_{\text{MA}} \leq
\sigma_{\text{E}}$, one can see that the meta-analytic standard error is never
larger than Edgington's standard error, with equality if and only if the trial
standard error coincide ($\sigma_1 = \sigma_2$). Similarly, by applying
algebraic manipulations to $\sigma_{\text{E}} \leq \sigma_1$ and
$\sigma_{\text{E}} \leq \sigma_2$, one can see that Edgington's standard error
is only equal or smaller than either trial standard error if $\sqrt{2} - 1 \leq
\sigma_2/\sigma_1 \leq 1/(\sqrt{2} - 1) = \sqrt{2} + 1$.

For the remaining methods, the (approximate) median estimates are given by $X$
and $Y$ in equation~\eqref{eq:XandY} and Table~\ref{tab:constant} with $a =
1/2$. We can therefore use the results from Nadarajah and Kotz
\citep{Nadarajah2008} to obtain their second moments
\begin{align*}
  E(X^2) =&
  \{\sigma^2_1 + (\theta_1 + \sigma_1\,q)^2\}\, \Phi\left(\frac{\theta_2 - \theta_1 + q(\sigma_2 - \sigma_1)}{\sqrt{\sigma^2_1 + \sigma^2_2}}\right) +
  \{\sigma^2_2 + (\theta_2 + \sigma_2\,q)^2\} \, \Phi\left(\frac{\theta_1 - \theta_2 + q(\sigma_1 - \sigma_2)}{\sqrt{\sigma^2_1 + \sigma^2_2}}\right) \\
  & -
  \{\theta_1 + \theta_2 + q(\sigma_1 + \sigma_2)\} \sqrt{\sigma_1^2 + \sigma_2^2} \, \phi\left(\frac{\theta_2 - \theta_1 + q(\sigma_2^2 - \sigma_1^2)}{\sqrt{\sigma_1 + \sigma_2}}\right)
\end{align*}
and
\begin{align*}
  E(Y^2) =&
  \{\sigma^2_1 + (\theta_1 - \sigma_1\,q)^2\}\, \Phi\left(\frac{\theta_1 - \theta_2 + q(\sigma_2 - \sigma_1)}{\sqrt{\sigma^2_1 + \sigma^2_2}}\right) +
  \{\sigma^2_2 + (\theta_2 - \sigma_2\,q)^2\} \, \Phi\left(\frac{\theta_2 - \theta_1 + q(\sigma_1 - \sigma_2)}{\sqrt{\sigma^2_1 + \sigma^2_2}}\right) \\
  & +
  \{\theta_1 + \theta_2 - q(\sigma_1 + \sigma_2)\} \sqrt{\sigma_1^2 + \sigma_2^2} \, \phi\left(\frac{\theta_1 - \theta_2 + q(\sigma_2 - \sigma_1)}{\sqrt{\sigma_1^2 + \sigma_2^2}}\right)
\end{align*}
and corresponding standard errors.
For example, assuming equal standard errors ($\sigma_1 = \sigma_2 = \sigma$),
equal true effects ($\theta_1 = \theta_2 = \theta$), and the alternative
``greater'', the standard error of the two-trials rule median estimate ($X$ with
$q = z_{\sqrt{1/2}}$) simplifies to~\eqref{eq:se2tr}.

\begin{figure}[!htb]
\begin{knitrout}
\definecolor{shadecolor}{rgb}{0.969, 0.969, 0.969}\color{fgcolor}
\includegraphics[width=\maxwidth]{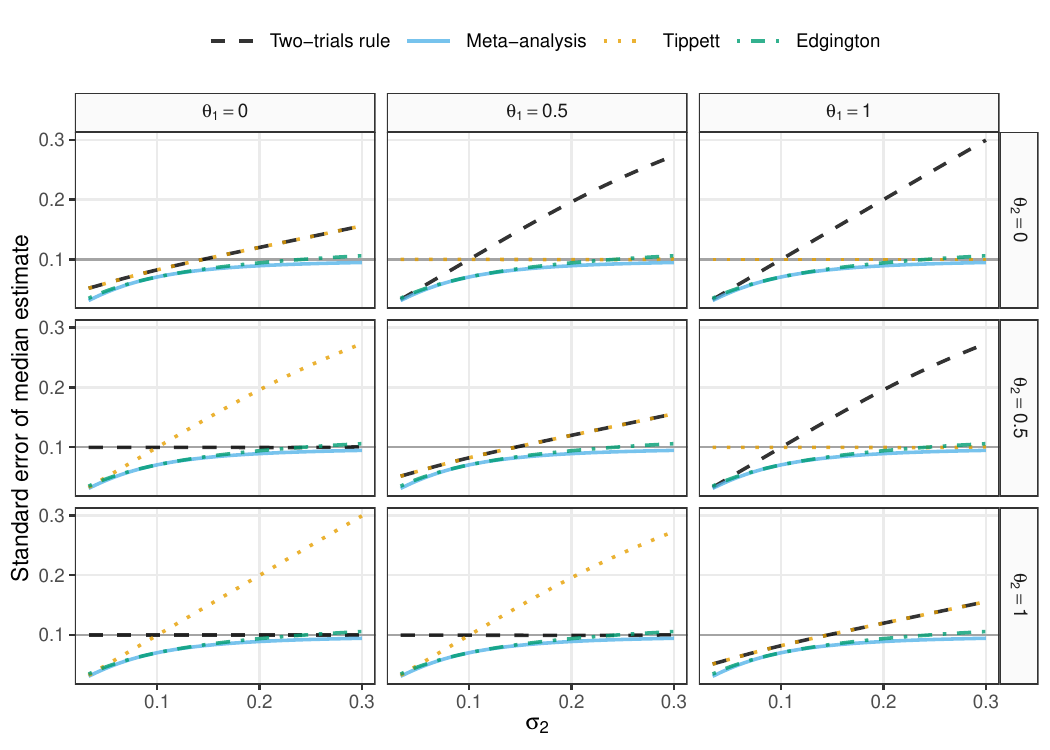} 
\end{knitrout}
\caption{Comparison of median estimate standard errors across different
  scenarios of true trial effects $\theta_1$ and $\theta_2$ and trial standard
  error from the second trial $\sigma_2$. The standard error of the first trial
  is $\sigma_1 = 0.1$ across all scenarios. The standard errors from
  Fisher's and Pearson's methods are close to Tippett's method and the
  two-trials rule, and not shown to make the plot easier to read.}
\label{fig:ses}
\end{figure}

Figure~\ref{fig:ses} shows the standard errors from the median estimates of the
two-trials rule, meta-analysis, Tippett's, and Edgington's methods for various
scenarios of true trial effects and trial standard errors that should resemble
typical ranges for standardized mean difference parameters. The approximate
standard errors from Fisher's and Pearson's methods are close to the ones from
Tippett's method and the two-trials rule, and therefore not shown to make the
plot easier to read.

We can see that the standard error from meta-analysis is always the lowest and
is always smaller than the minimum of the two standard errors
($\sigma_{\text{MA}} \leq \min\{\sigma_1, \sigma_2\}$). The standard error from
Edgington's method is equal (when $\sigma_1 = \sigma_2$) to or larger than the
meta-analytic one, and can exceed the minimum standard error from the two trials
(e.g., for $\sigma_1 = 0.1$ and $\sigma_2 = 0.3$, it is $\sigma_{\text{E}} =
0.106$ ). The standard errors for both Edgington's
method and meta-analysis only depend on the trials' standard errors, but not on
the true effects, so their standard errors are the same across all panels in
Figure~\ref{fig:ses}. This is not the case for the two-trials rule and Tippett's
method, which show a more irregular behavior. When, the true trial effects
coincide ($\theta_1 = \theta_2$; panels on the diagonal), the combined standard
error decreases with decreasing standard error from the second trial $\sigma_2$.
However, when the true effect from the first trial is smaller than the one from
the second trial ($\theta_1 < \theta_2$; upper off-diagonal panels), the
combined standard error from Tippett's method remains constant whereas the
standard error from the two-trials rule changes drastically with changing
$\sigma_2$. The opposite occurs when the effect from the first trial is larger
($\theta_1 > \theta_2$; lower off-diagonal panels). This is plausible, as these
methods target the minimum (two-trials rule) or maximum (Tippett) effect,
meaning that the standard error of the trial with minimum or maximum effect
mainly affects the standard error of the combined estimate.

\bmsubsection{Limiting combined estimation functions}
\label{app:asymptotics}

Consider again the random variables $X$ and $Y$ as defined in
Appendix~\ref{app:expectation}. Their cumulative distribution functions can be
derived to be
\begin{align*}
  \Pr(X \leq x)
  &= \Pr(\min\{\hat{\theta}_1 + \sigma_1 \, q, \hat{\theta}_2 + \sigma_2 \, q\} \leq x) \\
  &= 1 - \Pr(\hat{\theta}_1 + \sigma_1 \, q > x, \hat{\theta}_2 + \sigma_2 \, q > x) \\
  &= 1 - \{\Pr(\hat{\theta}_1 + \sigma_1 \, q > x) \times \Pr(\hat{\theta}_2 + \sigma_2 \, q > x)\} \\
  &= 1 - \left\{\Phi\left(\frac{\theta_1 - x}{\sigma_1} + q\right) \times \Phi\left(\frac{ \theta_2 - x}{\sigma_2} + q\right)\right\}.
\end{align*}
and
\begin{align*}
  \Pr(Y \leq y)
  &= \Pr(\max\{\hat{\theta}_1 - \sigma_1 \, q, \hat{\theta}_2 - \sigma_2 \, q\} \leq y) \\
  &= \Pr(\hat{\theta}_1 - \sigma_1 \, q \leq y, \hat{\theta}_2 - \sigma_2 \, q \leq y) \\
  &= \Pr(\hat{\theta}_1 - \sigma_1 \, q \leq y) \times \Pr(\hat{\theta}_2 - \sigma_2 \, q \leq y) \\
  &= \Phi\left(\frac{y - \theta_1}{\sigma_1} + q\right) \times \Phi\left(\frac{y - \theta_2}{\sigma_2} + q\right)
\end{align*}
Letting the standard errors $\sigma_1$ and $\sigma_2$ go to zero, this leads to
\begin{align}
  \lim_{\sigma_1, \sigma_2 \downarrow 0} \Pr(X \leq x)
  &= 1 - \{1_{(-\infty, \theta_1)}(x) \times 1_{(-\infty, \theta_2)}(x)\} \nonumber \\
  &= 1_{[\min\{\theta_1, \theta_2\}, +\infty)}(x)
  \label{eq:limitcdfx}
\end{align}
and
\begin{align}
  \label{eq:limitcdfy}
  \lim_{\sigma_1, \sigma_2 \downarrow 0} \Pr(Y \leq y)
  &= 1_{[\theta_1, +\infty)}(y) \times 1_{[\theta_2, +\infty)}(y) \nonumber \\
  &= 1_{[\max\{\theta_1, \theta_2\}, +\infty)}(y)
\end{align}
where $1_A(x) = 1$ if $x \in A$ and 0 otherwise, and the constant $q$ vanishes.
Since~\eqref{eq:limitcdfx} and~\eqref{eq:limitcdfy} is the cumulative
distribution function of a degenerate random variable at $\min\{\theta_1,
\theta_2\}$ and $\max\{\theta_1, \theta_2\}$, respectively, this implies that
the combined estimation functions given by $X$ and $Y$ converge in probability
to $\min\{\theta_1, \theta_2\}$ and $\max\{\theta_1, \theta_2\}$, respectively,
for any constant $q$. Thus, all combined estimation functions from
Table~\ref{tab:constant} converge in probability to $\min\{\theta_1, \theta_2\}$
or $\max\{\theta_1, \theta_2\}$ as $\sigma_1$ and $\sigma_2$ decrease.

\begin{figure}[!htb]
\begin{knitrout}
\definecolor{shadecolor}{rgb}{0.969, 0.969, 0.969}\color{fgcolor}
\includegraphics[width=\maxwidth]{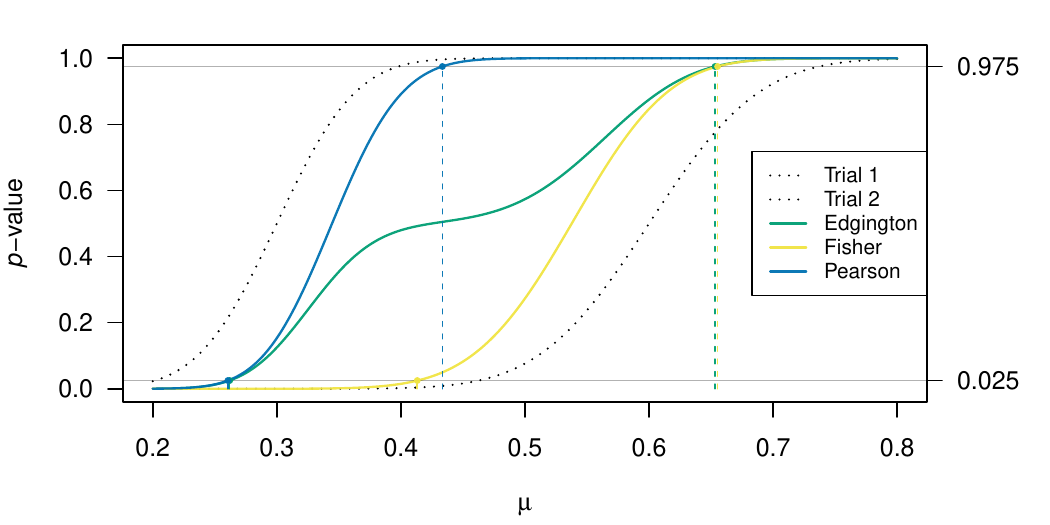} 
\end{knitrout}
\caption{Two well-separated \textit{p}-value functions (with alternative ``greater'')
  and the associated combined \textit{p}-value functions based on Fisher's, Pearson's,
  and Edgington's methods. The dashed vertical lines and points denote the 95\%
  CI limits computed with the approximate combined estimation
  functions~\eqref{eq:fisherest2}, \eqref{eq:pearsonest2},
  and~\eqref{eq:edgingtonprecise}. The effect estimates are $\hat{\theta}_1 =
  0.3$ and $\hat{\theta}_2 = 0.6$ while the standard errors are
  $\sigma_1 = 0.05$ and $\sigma_2 = 0.07$.}
\label{fig:separated-pvals}
\end{figure}

\bmsubsection{Approximate combined estimation functions}
\label{app:approxmu}
Suppose that the trials' individual \textit{p}-value functions
\begin{align*}
  &p_1(\mu) =
  \begin{cases}
    1 - \Phi\left(\frac{\hat{\theta}_1 - \mu}{\sigma_1}\right) & \text{for alternative = ``greater''} \\
     \Phi\left(\frac{\hat{\theta}_1 - \mu}{\sigma_1}\right) & \text{for alternative ``less''} \\
    \end{cases}&
  &~&
   &p_2(\mu) =
  \begin{cases}
    1 - \Phi\left(\frac{\hat{\theta}_2 - \mu}{\sigma_2}\right) & \text{for alternative = ``greater''} \\
     \Phi\left(\frac{\hat{\theta}_2 - \mu}{\sigma_2}\right) & \text{for alternative = ``less''} \\
    \end{cases}&
\end{align*}
are ``well-separated'' in the sense that in the region where $p_1(\mu)$ changes
from 0 to 1, $p_2(\mu)$ stays almost constant at 0 or 1, see the dotted lines in
Figure~\ref{fig:separated-pvals} for an example. This happens when either the
estimates $\hat{\theta}_1$ and $\hat{\theta}_2$ are far apart and/or the
standard errors $\sigma_1$ and $\sigma_2$ are small relative to the estimates
(provided the estimates are not equal). Note that asymptotically the individual
\textit{p}-value functions approach the step functions
\begin{align*}
  &\lim_{\sigma_1 \downarrow 0} p_1(\mu) =
  \begin{cases}
    1_{[\theta_1, +\infty)}(\mu) & \text{for alternative = ``greater''} \\
     1_{(-\infty, \theta_1]}(\mu) & \text{for alternative = ``less''} \\
    \end{cases}&
    &~&
  &\lim_{\sigma_2 \downarrow 0} p_2(\mu) =
    \begin{cases}
    1_{[\theta_2, +\infty)}(\mu) & \text{for alternative = ``greater''} \\
     1_{(-\infty, \theta_2]}(\mu) & \text{foralternative = ``less''} \\
    \end{cases}&
\end{align*}
Hence, with decreasing standard errors, the trials' \textit{p}-value functions
eventually become well-separated whenever the true effects $\theta_1$ and
$\theta_2$ are unequal.

In case of well-separated \textit{p}-value functions, we can approximate the combined
\textit{p}-value function from Fisher's, Pearson's, and Edgington's method by setting
one of the \textit{p}-values to 0 or 1, depending on alternative and combination
method, and derive an approximate but closed-form combined estimation function.
For example, in Figure~\ref{fig:separated-pvals} the combined \textit{p}-value from
Fisher's method~\eqref{eq:fisher} remains virtually constant for increasing
$\mu$ when the first individual \textit{p}-function increases and only starts to
increase as the second \textit{p}-value function increases.

We may hence approximate Fisher's combined \textit{p}-value by
\begin{align}
  p_{\text{F}}(\mu) =
  \begin{cases}
    1 - \Pr\left[\chi^{2}_{4} \leq -2 
    \log\left\{1 - \Phi\left(\max\left\{\frac{\hat{\theta}_1 - \mu}{\sigma_1}, \frac{\hat{\theta}_2 - \mu}{\sigma_2}\right\}\right)\right\} 
    \right] &  \text{for alternative = "greater"} \\
  1 - \Pr\left[\chi^{2}_{4} \leq -2 
    \log\left\{\Phi\left(\min\left\{\frac{\hat{\theta}_1 - \mu}{\sigma_1}, \frac{\hat{\theta}_2 - \mu}{\sigma_2}\right\}\right)\right\} 
    \right] &  \text{for alternative = "less".} \\
  \end{cases}
  \label{eq:fisherapprox}
\end{align}
The corresponding combined estimation function can then be obtained by
equating~\eqref{eq:fisherapprox} to $a$ and solving for $\mu$, which leads to
\begin{equation}
  \label{eq:fisherest2}
  \hat{\mu}_{\text{F}}(a) =
  \begin{cases}
    \max\{\hat{\theta}_{1} + \sigma_{1} \, z_{\exp\{-\chi^{2}_{4}(1 - a)/2\}},
    \hat{\theta}_{2} + \sigma_{2} \, z_{\exp\{-\chi^{2}_{4}(1 - a)/2\}}\} &  \text{for alternative = "greater"} \\
    \min\{\hat{\theta}_{1} - \sigma_{1} \, z_{\exp\{-\chi^{2}_{4}(1 - a)/2\}},
    \hat{\theta}_{2} - \sigma_{2} \, z_{\exp\{-\chi^{2}_{4}(1 - a)/2\}}\} &  \text{for alternative = "less".} \\
  \end{cases}
\end{equation}
The dashed yellow vertical lines in Figure~\ref{fig:separated-pvals} show the
limits of a 95\% CI computed via~\eqref{eq:fisherest2}, demonstrating that the
approximation is accurate in this case, despite the finite standard errors.

In an analogous fashion, the combined \textit{p}-value function based on Pearson's
method can be approximated by
\begin{align*}
  p_{\text{P}}(\mu) =
  \begin{cases}
    \Pr\left[\chi^{2}_{4} \leq -2 
      \log\left\{\Phi\left(\min\left\{\frac{\hat{\theta}_1 - \mu}{\sigma_1}, \frac{\hat{\theta}_2 - \mu}{\sigma_2}\right\}\right)\right\} 
    \right] &  \text{for alternative = "greater"} \\
      \Pr\left[\chi^{2}_{4} \leq -2 
      \log\left\{1 - \Phi\left(\max\left\{\frac{\hat{\theta}_1 - \mu}{\sigma_1}, \frac{\hat{\theta}_2 - \mu}{\sigma_2}\right\}\right)\right\} 
    \right] &  \text{for alternative = "less"}
  \end{cases}
\end{align*}
leading to the approximate combined estimation function
\begin{equation}
  \label{eq:pearsonest2}
  \hat{\mu}_{\text{P}}(a) =
  \begin{cases}
    \min\{\hat{\theta}_{1} - \sigma_{1} \, z_{\exp\{-\chi^{2}_{4}(a)/2\}},
    \hat{\theta}_{2} - \sigma_{2} \, z_{\exp\{-\chi^{2}_{4}(a)/2\}}\} &  \text{for alternative = "greater"} \\
    \max\{\hat{\theta}_{1} + \sigma_{1} \, z_{\exp\{-\chi^{2}_{4}(a)/2\}},
    \hat{\theta}_{2} + \sigma_{2} \, z_{\exp\{-\chi^{2}_{4}(a)/2\}}\} &  \text{for alternative = "less"}. \\
  \end{cases}
\end{equation}
The functions~\eqref{eq:fisherest2} and~\eqref{eq:pearsonest2} based on Fisher's
and Pearson's methods have a striking similarity to the combined estimation
functions based on Tippett's method~\eqref{eq:tippest} and the two-trials
rule~\eqref{eq:2trialsest}, respectively, as they again involve shifted
maxima/minima of the trial estimates.

In a similar way, Edgington's combined \textit{p}-value function can be approximated by
\begin{align*}
  p_{\text{E}}(\mu) =
  \begin{cases}
    \left\{1 - \Phi\left(\min\left\{\frac{\hat{\theta}_1 - \mu}{\sigma_1}, \frac{\hat{\theta}_2 - \mu}{\sigma_2}\right\}\right)\right\}^{2} \big/ \,2
    & \text{if} ~ \mu < \dfrac{\hat{\theta}_1/\sigma_1 + \hat{\theta}_2/\sigma_2}{1/\sigma_1 + 1/\sigma_2} \\
    \dfrac{1}{2} & \text{if} ~  \mu = \dfrac{\hat{\theta}_1/\sigma_1 + \hat{\theta}_2/\sigma_2}{1/\sigma_1 + 1/\sigma_2} \\
    1 - \left\{\Phi\left(\max\left\{\frac{\hat{\theta}_1 - \mu}{\sigma_1}, \frac{\hat{\theta}_2 - \mu}{\sigma_2}\right\}\right)\right\}^{2} \big/ \,2 & \text{else}
  \end{cases}
\end{align*}
for the alternative ``greater'' and with
\begin{align*}
  p_{\text{E}}(\mu) =
  \begin{cases}
    \left\{ \Phi\left(\max\left\{\frac{\hat{\theta}_1 - \mu}{\sigma_1}, \frac{\hat{\theta}_2 - \mu}{\sigma_2}\right\}\right)\right\}^{2} \big/ \,2
    & \text{if} ~  \mu > \dfrac{\hat{\theta}_1/\sigma_1 + \hat{\theta}_2/\sigma_2}{1/\sigma_1 + 1/\sigma_2}  \\
    \dfrac{1}{2} & \text{if} ~  \mu = \dfrac{\hat{\theta}_1/\sigma_1 + \hat{\theta}_2/\sigma_2}{1/\sigma_1 + 1/\sigma_2} \\
    1- \left\{1 - \Phi\left(\min\left\{\frac{\hat{\theta}_1 - \mu}{\sigma_1}, \frac{\hat{\theta}_2 - \mu}{\sigma_2}\right\}\right)\right\}^{2} \big/ \,2 & \text{else}
  \end{cases}
\end{align*}
for the alternative ``less''. Consequently, the approximate combined estimation
function is
\begin{equation}
  \label{eq:edgingtonprecise}
  \hat{\mu}_{\text{E}}(a) =
  \begin{cases}
    \min\{\hat{\theta}_{1} + \sigma_{1} \, z_{\sqrt{2a}}, \hat{\theta}_{2} + \sigma_{2} \, z_{\sqrt{2a}}\} & \text{for} ~ a < 1/2 \\
    \dfrac{\hat{\theta}_1/\sigma_1 + \hat{\theta}_2/\sigma_2}{1/\sigma_1 + 1/\sigma_2} & \text{for} ~ a = 1/2 \\
    \max\{\hat{\theta}_{1} - \sigma_{1} \, z_{\sqrt{2(1 - a)}}, \hat{\theta}_{2} - \sigma_{2} \, z_{\sqrt{2(1 - a)}}\} & \text{for} ~ a > 1/2 \\
  \end{cases}
\end{equation}
for the alternative ``greater'' and
\begin{equation}
  \label{eq:edgingtonprecise2}
  \hat{\mu}_{\text{E}}(a) =
  \begin{cases}
    \max\{\hat{\theta}_{1} - \sigma_{1} \, z_{\sqrt{2a}}, \hat{\theta}_{2} - \sigma_{2} \, z_{\sqrt{2a}}\} & \text{for} ~ a < 1/2 \\
    \dfrac{\hat{\theta}_1/\sigma_1 + \hat{\theta}_2/\sigma_2}{1/\sigma_1 + 1/\sigma_2} & \text{for} ~ a = 1/2 \\
    \min\{\hat{\theta}_{1} + \sigma_{1} \, z_{\sqrt{2(1 - a)}}, \hat{\theta}_{2} + \sigma_{2} \, z_{\sqrt{2(1 - a)}}\} & \text{for} ~ a > 1/2 \\
  \end{cases}
\end{equation}
for the alternative ``less''. The combined estimation
functions~\eqref{eq:edgingtonprecise} and~\eqref{eq:edgingtonprecise2} also
include the closed-form solution for the median estimate ($a = 1/2$)
from~\eqref{eq:edgingtonpointest} as this value does not require any
approximation. Surprisingly, a $(1 - \alpha) \times 100\%$ CI with $\alpha <
1/4$ constructed from~\eqref{eq:edgingtonprecise}
or~\eqref{eq:edgingtonprecise2} always includes the individual estimates
$\hat{\theta}_1$ and $\hat{\theta}_2$ since the lower limit ($a = \alpha/2$) is
always smaller than the minimum of the two effect estimates, and the upper limit
($a = 1 - \alpha/2$) is always larger than the maximum of the two. This
demonstrates that Edgington's method reacts to heterogeneity by widening its CI
to include both trial effect estimates.

All of these approximations become more accurate with decreasing standard errors
as the individual \textit{p}-value functions become more separated. Since all
approximate combined estimation
functions~\eqref{eq:fisherest2}--\eqref{eq:edgingtonprecise2} are essentially
shifted minima and maxima (apart from the median estimate of Edgington's
method), the results from Appendix~\ref{app:asymptotics} apply. That is, as the
standard errors $\sigma_1$ and $\sigma_2$ decrease toward zero, all minima
converge in probability to $\min\{\theta_1,\theta_2\}$ while all maxima converge
to $\max\{\theta_1,\theta_2\}$.

\end{document}